\documentclass[sigconf]{acmart}
\copyrightyear{2024} 
\acmYear{2024} 
\setcopyright{rightsretained}
\acmConference[WWW '24] {Proceedings of the ACM Web Conference 2024}{May 13--17, 2024}{Singapore, Singapore.}
\acmBooktitle{Proceedings of the ACM Web Conference 2024 (WWW '24), May 13--17, 2024, Singapore, Singapore}
\acmISBN{979-8-4007-0171-9/24/05} 
\acmDOI{10.1145/3589334.3645649}

\usepackage[ruled,vlined]{algorithm2e}
\usepackage{multirow}
\usepackage{booktabs,caption}
\usepackage[flushleft]{threeparttable}

\usepackage{latexsym}
\usepackage{amsmath}

\begin{document}

%%
%% The "title" command has an optional parameter,
%% allowing the author to define a "short title" to be used in page headers.
%\title{Heterogeneous Hypergraph for Multi-Label Zero-shot Product Attribute-Value Extraction}
\title{Multi-Label Zero-Shot Product Attribute-Value Extraction}

%%
%% The "author" command and its associated commands are used to define
%% the authors and their affiliations.
%% Of note is the shared affiliation of the first two authors, and the
%% "authornote" and "authornotemark" commands
%% used to denote shared contribution to the research.
\author{Jiaying Gong}
\affiliation{%
  \institution{Virginia Tech}
  \city{Blacksburg}
  \country{U.S.}}
\email{gjiaying@vt.edu}

\author{Hoda Eldardiry}
\affiliation{%
  \institution{Virginia Tech}
  \city{Blacksburg}
  \country{U.S.}}
\email{hdardiry@vt.edu}

%%
%% By default, the full list of authors will be used in the page
%% headers. Often, this list is too long, and will overlap
%% other information printed in the page headers. This command allows
%% the author to define a more concise list
%% of authors' names for this purpose.
\renewcommand{\shortauthors}{Jiaying Gong and Hoda Eldardiry}

%%
%% The abstract is a short summary of the work to be presented in the
%% article.

\begin{abstract}
%For the success of e-commerce platforms, it is critical to provide attribute values from the product profile for better search and recommendation.
%However, attribute value information can be unavailable for new products. 
%It is hard for traditional supervised learning models to extract unseen attribute values as they require large quantities of labeled training data.
%It is difficult, time-consuming, and costly to manually label large quantities of new product profiles.
%To identify missing labels, traditional deep learning methods rely on a large amount of labeled data. It is difficult, time-consuming and costly to manually label large quantities of new product profiles for training. 
E-commerce platforms should provide detailed product descriptions (attribute values) for effective product search and recommendation. However, attribute value information is typically not available for new products. To predict unseen attribute values, large quantities of labeled training data are needed to train a traditional supervised learning model. Typically, it is difficult, time-consuming, and costly to manually label large quantities of new product profiles.
%To efficiently extract unseen attribute values from new products, we propose a heterogeneous hypergraph with an inductive ability for zero-shot (unseen) product attribute value extraction (HyperPAVE). 
%By constructing heterogeneous hypergraphs, complex higher-order relations (i.e. user behavior information) are captured when updating the node features. 
%With an inductive link prediction mechanism to infer future connections between unseen nodes, new attribute values can be identified without labeled training data.
%The key advantage of HyperPAVE is that it can be trained to capture the underlying hidden patterns, relationships, and features. 
%This enables the model to generalize well to predict new, unseen nodes and edges. 
%Extensive experiments with ablation studies conducted on different categories of public dataset MAVE demonstrate that our proposed model significantly outperforms the existing classification-based and generation-based models for attribute value extraction in the zero-shot setting. 
In this paper, we propose a novel method to efficiently and effectively extract unseen attribute values from new products in the absence of labeled data (zero-shot setting). We propose HyperPAVE, a multi-label zero-shot attribute value extraction model that leverages inductive inference in heterogeneous hypergraphs. %an inductive link prediction mechanism in heterogeneous hypergraphs.
%HyperPAVE, heterogeneous Hypergraph with an inductive ability for zero-shot (unseen) product attribute value extraction. 
In particular, our proposed technique constructs heterogeneous hypergraphs to capture complex higher-order relations (i.e. user behavior information) to learn more accurate feature representations for graph nodes. Furthermore, our proposed HyperPAVE model uses an inductive link prediction mechanism to infer future connections between unseen nodes. This enables HyperPAVE to identify new attribute values without the need for labeled training data.
%The key advantage of HyperPAVE is that it can be trained to capture the underlying hidden patterns, relationships, and features. 
%This enables the model to generalize well to predict new, unseen nodes and edges. 
We conduct extensive experiments with ablation studies on different categories of the MAVE dataset. The results demonstrate that our proposed HyperPAVE model significantly outperforms existing classification-based, generation-based large language models for attribute value extraction in the zero-shot setting. 
%The code will be released after acceptance.
\end{abstract}

%%
%% The code below is generated by the tool at http://dl.acm.org/ccs.cfm.
%% Please copy and paste the code instead of the example below.
%%


%\ccsdesc[500]{Computer systems organization~Embedded systems}
%\ccsdesc[300]{Computer systems organization~Redundancy}
%\ccsdesc{Computer systems organization~Robotics}
%\ccsdesc[100]{Networks~Network reliability}

%%
%% Keywords. The author(s) should pick words that accurately describe
%% the work being presented. Separate the keywords with commas.

\maketitle
%\vspace{-10mm}
\section{Introduction}
% AVE Introduction
Product attribute value extraction (AVE) aims to extract attribute-value pairs (i.e. <color: red>) from e-Commerce product descriptions, which provides a better search and recommendation experience for customers.
Existing studies on AVE mainly focus on supervised-learning models such as sequence labeling~\cite{yan-etal-2021-adatag, jain-etal-2021-learning}, extractive question answering~\cite{wang2020learning, shinzato2022simple} and multi-modal learning~\cite{wang2022smartave, liu2023boosting, ghosh2023d, wang2023mpkgac} models.
These supervised learning models are trained to only predict seen attribute value pairs.
However, new products with unseen attribute-value pairs enter the market every day in real-world e-commerce platforms. 
It is time-consuming and costly to manually label large quantities of new products for training. 

Some recent works focus on open mining models~\cite{10.1145/3485447.3512035, xu2023towards} to directly extract attribute values from product titles or descriptions.
However, these approaches can not discover attribute values that are not explicitly mentioned in the text.
In other words, these open mining models can not extract values that never appear in the product profile.
To extract unseen attribute values, these open mining models use self-supervised learning, but they still need a high-quality seed attribute set bootstrapped from existing resources.
Besides these open mining models, some generative large language models (LLM) are fine-tuned to autoregressively decode unseen attribute values from the input text. 
However, fine-tuning such LLM (i.e. T5~\cite{raffel2020exploring}) requires a lot of time and computing resources.

\begin{figure}[htp] 
 \center{\includegraphics[height=6.2cm,width=8.2cm]{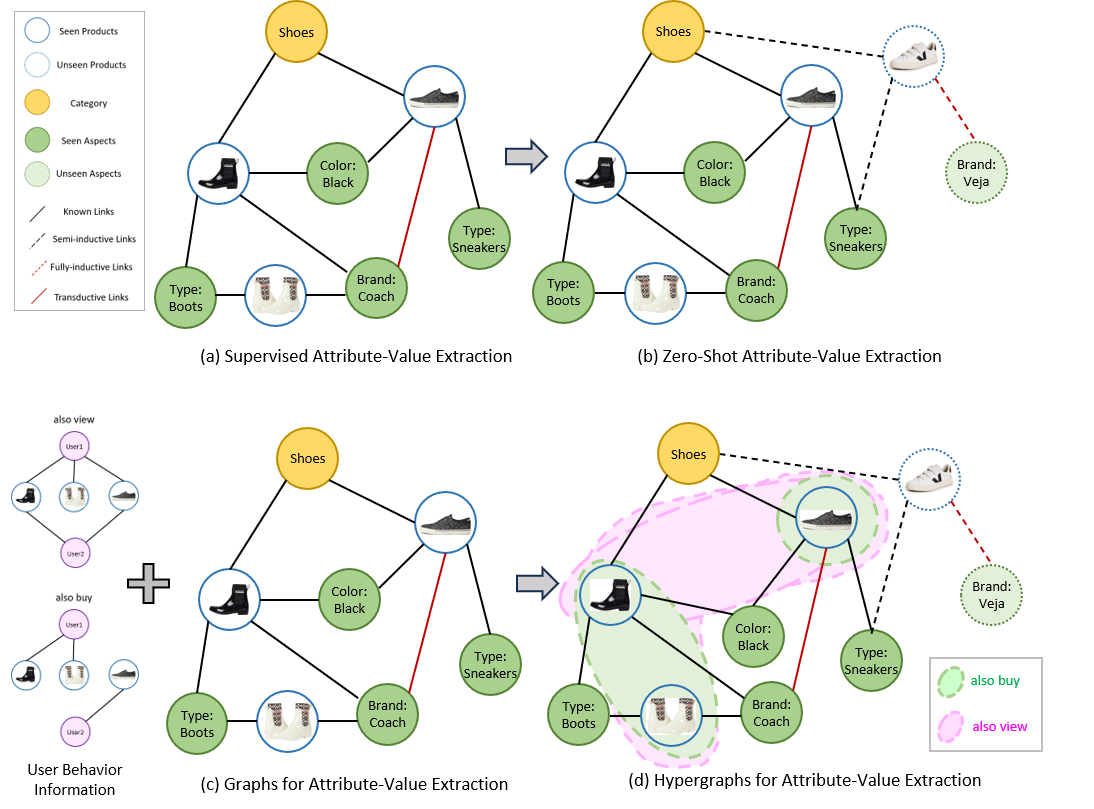}}
 \caption{\label{fig:example} An example of zero-shot product attribute-value extraction by semi-inductive link predictions.}
 \end{figure}

%Solutions
% First motivated by inductive linke prediction, zero-shot. Second, higher-order
To address the above challenges, we propose HyperPAVE, a multi-label zero-shot attribute value extraction model that leverages inductive inference in heterogeneous hypergraphs to recognize unseen (new) attribute-value pairs (aspects) for which there is no available labeled training data. 
%Motivation
Motivated by the inductive graph learning, which shows the superiority of GNN to inductively adapt to infer unseen nodes~\cite{fan2023zero, xu2020inductive}, we build inductive heterogeneous hypergraphs employing inductive link prediction mechanisms to infer missing or future connections (e.g., from new ‘product’ node to unseen ‘aspect’ node). 
%where the product graph is updated with zero-shot product and attribute values. 
%Then, message-passing is conducted directly on the updated product graph, ensuring the inductive inference ability.
The top part of Figure 1 shows an example comparison between supervised (Figure 1a) and zero-shot (Figure 1b) attribute value extraction.
Existing works formulate relation propagation as a transductive link prediction task (Figure 1a), where links can only be predicted between seen nodes (products and aspects)~\cite{ma2020transductive, chen2020hgmf}. 
To recognize unseen (new) aspects for new products, negative links are added in the original graph and the model is trained to predict whether an edge exists between two nodes based on the node features. 
HyperPAVE aims to learn the connections between both the nodes’ features that are obtained from the fine-tuned LLM-based encoder and the complex graph structure.
Motivated by the success of combining inductive GNNs and pre-trained BERT models~\cite{huang2022contexting}, HyperPAVE is designed to enhance the inductive hypergraph-based model with fine-tuned BERT contextual embeddings for each node.
Then, HyperPAVE is updated with zero-shot products and aspects with fine-tuned contextual embeddings, where message-passing is conducted directly on the updated graph, ensuring the inductive inference ability.

In addition, given the complexity of product data, it is important to design a model that can capture the heterogeneous, interconnected, and higher-order representation of both product data and user behavior data. 
Therefore, our proposed model HyperPAVE consists of various types of nodes including ‘category’, ‘product’, and ‘aspect’. 
%Node has information
The product node records information including both product titles and descriptions. 
To fully express the semantic information for attribute-value pairs, the aspect nodes record detailed attribute-value descriptions generated by a generator.
%) By leveraging both the label description generated by a generator and the category information as the auxiliary information to obtain more discriminative prototype.  To emphasize the difference between prototypes and reduce such ambiguity, we leverage more detailed label description generated by GPT-2 [ 20] to fully express the semantic information for attribute-value pairs and help learn more representative prototypes.
The proposed hypergraph representation uses higher-order relations to capture complex and interconnected user behavior information (e.g., `also buy', `also view') and product inventory information (e.g., `product has aspects', `category includes products').
The bottom part of Figure 1 shows an example comparison between graph-based (Figure 1c) and hypergraph-based (Figure 1d) attribute value extraction. To capture complex interconnected user behavior information, instead of using multiple graphs (one for each behavior e.g., “also buy” and “also view”), we construct hypergraphs using hyperedges to represent user behavior information as higher-order relations. 
Compared to using several different graphs to capture complex relations, using a hypergraph (1) can include more (i.e. user behavior) information for the final node representation, (2) does not need to include user nodes in the graph, and (3) relations are not limited to binary connections.
%The key advantage of HyperPAVE is that it can be trained to capture the underlying hidden patterns, relationships, and features. This enables the model to generalize well to predict new, unseen nodes and edges. 
The contributions are summarized as:
\begin{itemize}
%Two poinst: from structure information, we use heterogeneous hypergraphs, second, from node attribute features, we fine-tuned LLM
%we combine LLM with graphs to do inductive
    \item We propose a multi-label zero-shot model HyperPAVE to extract unseen attribute values for new products without labeled training data. HyperPAVE leverages an inductive link prediction mechanism combined with a fine-tuned BERT encoder to obtain unseen contextual node features.
   % HyperPAVE which employs an inductive link prediction mechanism to extract unseen attribute value pairs for new products without labeled training data.
    \item We build heterogeneous hypergraphs with higher-order relations to capture the complex and interconnected user behavior and structured product inventory information.% to capture the complex higher-order relations such as user behavior and product inventory information. % without using
    \item Extensive experiments on the public dataset MAVE demonstrate that HyperPAVE significantly outperforms the classification model, generative LLMs, and graph-based models in zero-shot learning. Besides, HyperPAVE also shows the effectiveness and efficiency of training.
    
   % to verify the effectiveness and efficiency of HyperPAVE for attribuve-value extraction in the zero-shot setting. The experimental results demonstrate that 
    %demonstrate that HyperPAVE significantly outperforms other classification and generation-based models for zero-shot attribute-value extraction on several categories of the public dataset MAVE. %Results also show the efficiency of HyperPAVE. we leverage a heterogenous hypergraph representations to include, higher order.....
\end{itemize}

%Contributions

\section{Related Works}
\subsection{Attribute Value Extraction}
Attribute value extraction (AVE) aims at extracting attribute-value pairs (aspects) based on the product information.
Early works use rule-based methods with domain-specific dictionaries to match target attribute value pairs~\cite{10.5555/2145432.2145598, 34460, 10.1145/1147234.1147241}.
With the development of neural networks, some studies view AVE as a sequence labeling problem~\cite{8731553, 10.1145/3219819.3219839, yan-etal-2021-adatag, jain-etal-2021-learning}.
Then, question-answering-based models are built to treat attributes as questions and values as answers~\cite{xu2019scaling, wang2020learning, shinzato2022simple}.
Multimodal fusion utilizing product images as visual features are learned to integrate visual semantics for products~\cite{zhu-etal-2020-multimodal, lin2021pam, wang2022smartave, cui2023pv2tea, liu2023boosting, ghosh2023d, wang2023mpkgac, liu2023multimodal}.
Some studies formulate AVE as a multi-label classification task to extract multiple aspects for the products~\cite{chen2022extreme, deng2022ae, gong2023knowledge}.
To handle unseen attribute values, open mining models~\cite{10.1145/3485447.3512035, xu2023towards} extract aspects directly from the text with limited/weak supervision, and generation models~\cite{shinzato2023unified} decode aspects as target sequences.
However, all of these approaches (1) require large quantities of labeled data for training and (2) miss higher-order relations between products, such as `also buy' or `also view' products.

\subsection{Zero-shot Learning}
Zero-shot learning has been widely applied in the field of computer vision (CV)~\cite{9832795} and natural language processing (NLP)~\cite{ijcai2021p597}.
Existing works for zero-shot learning in information extraction can be roughly divided into three categories: (1) Embedding-based models, where representations of both seen and unseen classes are learned based on the auxiliary information such as class information~\cite{chen-li-2021-zs, sainz-etal-2021-label} and other external information~\cite{10.1145/3459637.3482403, liu-etal-2022-pre}.
%Then, distances for representations of both seen classes and unseen classes are calculated, to determine the final prediction.
However, high-quality external knowledge is required for training the model, resulting in an increase in training time and resources.
(2) Generative-based models, where augmented samples are generated for unseen classes by generation models (i.e. GAN~\cite{mirza2014conditional}, VAEs~\cite{kingma2013auto}, and GPT-2~\cite{radford2019language}) based on the samples of seen classes. 
Then, the zero-shot learning problem is converted into a conventional supervised learning problem~\cite{chia-etal-2022-relationprompt, gong2023promptbased}.
However, these models suffer from the noise of augmented samples and performance highly depends on generative models.
(3) Graph-based models, where GNNs~\cite{scarselli2008graph} are directly used to predict unseen classes by inductive link prediction~\cite{ijcai2021p597}.
Most studies view this problem as zero-shot knowledge graph completion~\cite{geng2023benchmarking} or zero-shot item recommendation~\cite{fan2023zero}.
Attentive GCN is used to transfer features from seen classes to unseen classes~\cite{10.3233/SW-210435}.
Ontologies or topologies are utilized to augment ZSL by capturing relationships between classes~\cite{geng2022disentangled, chen2023integrating}.
Motivated by this, we build a product heterogeneous hypergraph to identify unseen aspects with inductive inference ability while capturing higher-order relations.

\subsection{Heterogeneous Hypergraph}
Hypergraphs are generalizations and extensions of ordinary graphs, where hyperedges can accommodate an arbitrary number of nodes to capture the higher-order relations~\cite{zhang2022hypergraph}.
To handle different types of nodes and edges, heterogeneous hypergraphs are learned by attention mechanisms~\cite{liu2023meta, li2023hypergraph, khan2023heterogeneous, ding-etal-2020-less}, wavelets~\cite{sun2021heterogeneous}, and variational auto-encoder~\cite{liu2022hypergraph, fan2021heterogeneous}.
Though, all of these works are widely applied for social networks~\cite{wei2023dual, li2022hmgcl}, academic citations~\cite{wu2022hypergraph, zhang2022learnable, 10010428}, biological networks~\cite{guzzi2022editorial, milano2022challenges} or product recommendation in e-commerce~\cite{xu2023correlative, cheng2022ihgnn, liu2023JDsearch, 10.1145/3539597.3570484}, heterogeneous hypergraphs are never applied to attribute value extraction in e-commerce. 
Different from the above hypergraphs that build hyperedges by close neighbors or meta-paths, we construct e-commerce related hyperedges by using user behavior and product inventory data to capture higher-order relations among categories, products, and aspects, to recognize unseen attribute values for new products.

\section{Methodology}
\begin{figure*}[htp] 
 \center{\includegraphics[height=10.5cm,width=\textwidth]{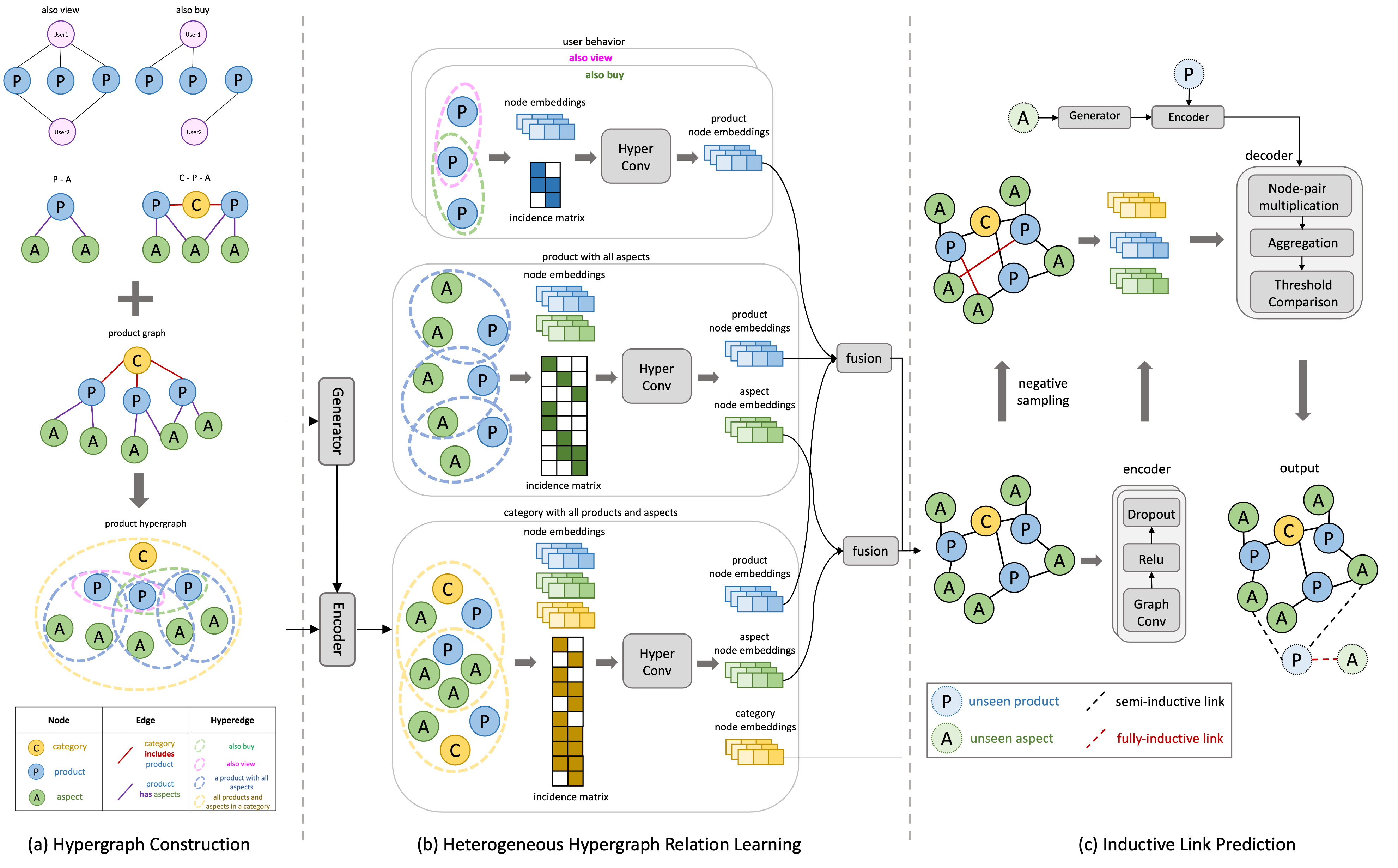}}
 \caption{\label{fig:framework} Overall framework of our proposed model HyperPAVE. The framework includes three key components: (a) Hypergraph Construction (b) Heterogeneous Hypergraph Relation Learning and (c) Inductive Link Prediction.}
 \end{figure*}

\subsection{Problem Definition}
In this section, we introduce the problem statement and some necessary definitions and notations for heterogeneous hypergraphs and multi-label zero-shot learning.

\paragraph{Problem Statement.} 
%TODO: different records, several graphs
Let $D=\left \{ c_{i}, p_{i}, a_{i} \right \}$ denote a corpus of e-commerce product records, where $c_{i}$, $p_{i}$, $a_{i}$ represent sub-category, product and attribute value pair (aspect), respectively. We use $C$, $P$, and $A$ to denote the sets of sub-categories, products, and aspects.
Hence, the task of attribute value extraction can be formulated as follows:
% We use U,L,T,A to denote the sets of users, locations, time-slots and activities respectively, and use NU , NL , NT , NA to represent the size of these sets. 
Input: The product records $D$.
Output: A model to estimate the probability that a new product $p$ in sub-category $c$ will have the unseen attribute value $a$. 
The goal of attribute value extraction is to learn a model $\mathcal{M}(p_{i}, a_{j})\rightarrow \hat{y}\left [ 0,1 \right ]$ to score the probability that a product $p_{i}$ has the attribute value $a_{j}$ based on $\mathcal{G}$, which includes all the relations from user behavior and product inventory information.
Given several different graphs (i.e. user behavior graphs, product inventory graphs, etc.), we first build a heterogeneous hypergraph $\mathcal{G}$ to capture the higher-order and non-binary relations contained in $\mathcal{G}$. 
Then, we aim to learn the representations for nodes on a heterogeneous hypergraph $\mathcal{G}$ for an inductive link prediction task.

\paragraph{DEFINITION 1 (Heterogeneous Hypergraph):} A heterogeneous hypergraph can be defined as $\mathcal{G} = \left \{ \mathcal{V}, \mathcal{E}, \mathcal{T}_{v}, \mathcal{T}_{e}, W \right \}$, where $\mathcal{V} = \left \{ v_{1}, v_{2}, \cdots, v_{N} \right \}$ is the node set, and $\mathcal{T}_{v}$ is the node type set. $\mathcal{E} = \left \{ e_{1}, e_{2}, \cdots, e_{M} \right \}$ is the hyperedge set, and $\mathcal{T}_{e}$ is the hyperedge type set, where $|\mathcal{T}_{v}| + |\mathcal{T}_{e}| > 2$.
$N$ and $M$ represent the maximum numbers of hyperedge nodes and edges.
$W = diag(w_{e_{1}}, w_{e_{2}}, \cdots, w_{e_{M}})$ denotes the diagonal matrix representing the hyperedge weight.
We use incidence matrix $H\in \mathbb{R}^{\left | \mathcal{V} \right | \times \left | \mathcal{E} \right | }$ to represent relationships between nodes and hyperedges, with entries defined as:
\begin{equation}
H(v, e) = \begin{cases}
1 & \text{ if } v \in e \\ 
0 & \text{ if } otherwise.
\end{cases}
\end{equation}
$D_{v} \in \mathbb{R}^{\left | \mathcal{V} \right | \times \left | \mathcal{V} \right |}$ and $D_{e} \in \mathbb{R}^{\left | \mathcal{E} \right | \times \left | \mathcal{E} \right |}$ are the diagonal matrices representing the degree matrix of nodes and hyperedges, where $D_{v}(i, i)=\sum_{e\in \mathcal{E}}^{}W(e)H(i,e)$ and $D_{e}(i, i)=\sum_{v\in \mathcal{V}}^{}H(v, i)$.
The normalized hypergraph adjacency matrix $A \in \mathbb{R^{\mathcal{V} \times \mathcal{V}}}$, representing the connection relationship between nodes, is defined as:
\begin{equation}
    A=D_{v}^{-1/2}HWD_{e}^{-1}H^{T}D_{v}^{-1/2}
\end{equation}

\paragraph{DEFINITION 2 (Zero-Shot Learning in Graph):} For multi-label zero-shot attribute-value (aspect) prediction, let $A^{s} = \left \{a_{1}^{s}, \cdots, a_{m}^{s} \right \}$ and $A^{u} = \left \{a_{1}^{u}, \cdots, a_{m}^{u} \right \}$ denote the node sets of seen and unseen aspects, where $A^{s}\cap A^{u} = \emptyset$. Only $A^{s}$ is included in the training graph $\mathcal{G}_{tr}$ and only $A^{u}$ is included in the testing graph $\mathcal{G}_{t}$. Product $p_{i}$ with any $a_{i}^{u}$ will be removed from $\mathcal{G}_{tr}$ to $\mathcal{G}_{t}$, to ensure all unseen aspect nodes are not in the training graph $\mathcal{G}_{tr}$. Details for multi-label zero-shot sampling are introduced in Algorithm~\ref{alg:data_sampling}.

\subsection{Multi-Label Zero-Shot Data Sampling}~\label{sec:zero_sampling}
Multi-label zero-shot data sampling includes (1) data splitting to ensure that there is \textbf{no overlap} of aspect and product nodes in training and validation/testing sets, and (2) negative sampling to balance the dataset.
For data splitting, we first randomly generate $N$ aspect nodes $A_{N}$ as unseen attribute values.
Then, we remove both the nodes $A_{N}$ and their corresponding products $P_{M}$ as unseen products, and all edges on $A_{N}$ and $P_{M}$ from the original graph $\mathcal{G}$, where $N \neq M$.
This step ensures that the zero-shot products and attribute values are never shown in the training graph.
We update the validation and testing graphs with the zero-shot nodes and links separately so that there's no overlap of zero-shot nodes and links between the validation and testing sets.
To balance the dataset, we do negative sampling and add negative links for all training, validation, and testing graphs.
Details for multi-label zero-shot data sampling are shown in Algorithm~\ref{alg:data_sampling}.

\begin{algorithm}[!htbp]
\SetKwInOut{KIN}{Input}
\SetKwInOut{KOUT}{Output}
\caption{Multi-label Zero-shot Data Sampling}
\label{alg:data_sampling} 
\KIN{Graph $\mathcal{G}$ with categories nodes $C$, product nodes $P$ and aspect nodes $A$, unseen aspect number $N$}
\KOUT{Train graph $\mathcal{G}_{tr}$, val graph $\mathcal{G}_{v}$, test graph $\mathcal{G}_{t}$}
Initialize $\mathcal{G}_{tr}$,  $\mathcal{G}_{v}$, $\mathcal{G}_{t}$\\
\For{$i$ in $Random(N)$}
{
$P_{i} = get\_node(\mathcal{G}, A_{i})$\\
$link_{pos}=get\_edge(\mathcal{G}, P_{i}, A_{i})$\\
$link_{neg}=Sampling(get\_complement(link_{pos}))$\\
$\mathcal{G}$.remove($A_{i}, P_{i}, link_{pos}$)\\
\eIf{$i$//2=0}
{
$\mathcal{G}_{v}$.update($A_{i}, P_{i}, link_{pos}, link_{neg}$)
}
{
$\mathcal{G}_{t}$.update($A_{i}, P_{i}, link_{pos}, link_{neg}$)
}
$\mathcal{G}_{tr} = \mathcal{G}.add\_negatives()$
}
return $\mathcal{G}_{tr}$, $\mathcal{G}_{v}$, $\mathcal{G}_{t}$
\end{algorithm}

\subsection{Overall Framework}
Figure~\ref{fig:framework} shows our proposed framework HyperPAVE with three main components: a) hypergraph construction, b) heterogeneous hypergraph relation learning, and c) inductive link prediction. We introduce each component in detail below.

\subsubsection{Heterogeneous Hypergraph Construction}~\label{sec:construction}
As shown in Figure~\ref{fig:framework}(a), there are three types of nodes: categories, products, and attribute values (aspects), and four types of hyperedges: `also view', `also buy', `product with all aspects' and `category with all products and aspects', which are constructed from two main data sources: user behavior information and product inventory information as:

\textit{1) User Behavior Data.} User behaviors have multiple types related to item-to-item relationships: people who bought X also bought Y (`also buy') and people who viewed X also viewed Y (`also view'). 
To well handle different user behaviors, we construct two types of hyperedges $\mathcal{T}_{e}^{u} = \left \{ \mathcal{E} ^{V} , \mathcal{E} ^{B} \right \}$, where $\mathcal{E} ^{V}$ represents `also view' and $\mathcal{E} ^{B}$ represents `also buy'. 
For example, given the record of user1 in `also view' graph shown in Figure~\ref{fig:framework}(a), we construct a hyperedge $\mathcal{E} ^{V}_{i}=\left \{ p_{1}, p_{2}, \cdots, p_{n} \right \} \in \mathcal{E}^{V}$ to model the interactions between users and products.
That is, each hyperedge in $\mathcal{E} ^{V}$ corresponds to one user.
These hyperedges are homogeneous because all nodes represent products.

\textit{2) Product Inventory Data.} Product inventory data refers to the existing product information records, including category, product, attribute values, etc.
We construct hyperedges $\mathcal{E} ^{P}$ to connect all attribute values to one product (P-A) and hyperedges $\mathcal{E}^{C}$ to connect all product information to one sub-category (C-P-A).
For example, given a product $p_{i}$, we construct a hyperedge $\mathcal{E}^{P}_{i}=\left \{ p_{i}, a_{1}, a_{2}, \cdots, a_{n} \right \} \in \mathcal{E}^{P}$ to indicate the relationships between product and its attribute values.
These heterogeneous hyperedges record the non-binary relations among categories, products, and attribute values.
To summarize it, we obtain hyperedge sets as:
\begin{equation}~\label{equ:hyperedge}
    \mathcal{T}_{e} = \left \{ \mathcal{E} ^{V} , \mathcal{E} ^{B}, \mathcal{E} ^{P}, \mathcal{E}^{C}\right \}
\end{equation}

\subsubsection{Heterogeneous Hypergraph Relation Learning}
\paragraph{Embedding Module.} As shown in Figure~\ref{fig:framework}(b), a heterogeneous hypergraph encoder first initializes the node embeddings. 
Since the attribute values (aspects) may lose contextual information due to the simple format, GPT-2~\cite{radford2019language} is adopted as the text generator to generate more detailed descriptions for attribute values.
For example, the attribute value: `connectivity: wireless' can be elaborated to a more detailed explanation: `connectivity is wireless communication between the user's device, which has an independent, physical signal to the user'.
%Encoder
We then adopt a pre-trained language model BERT~\cite{devlin-etal-2019-bert} as all nodes' input encoder to generate the initial contextual representation. 
For the product node, we construct a string [CLS;$t$;SEP;$d$] by concatenating the product title and description as the input, where CLS and SEP are special tokens.
The initial output representation for the category node $c_{i}$, product node $p_{i}$ and aspect node $a_{i}$ can be formulated as follows:

\begin{equation}~\label{equ:encoder}
h_{v_{c_{i}}} = tanh(W \cdot f_{\varnothing }(c_{i})+b)
\end{equation}
\begin{equation}~\label{equ:encoder}
h_{v_{p_{i}}} = tanh(W \cdot f_{\varnothing }(t_{i},d_{i})+b)
\end{equation}
\begin{equation}~\label{equ:label}
h_{v_{a_i}} = tanh(W \cdot f_{\varnothing }(g_{\varnothing}(a_{i}))+b)
\end{equation}
where $f_{\varnothing }$ is BERT encoder, $g_{\varnothing }$ is GPT-2 generator, $c$ is category, $t$ is product title, $d$ is product description, $a$ is `attribute value', $W$ and $b$ are trainable weights and bias.
To simplify the notations, we use $h_{v_i}$ to denote the initial feature embeddings of all different nodes.

\paragraph{Message Passing Module.}

\begin{table*}[]
\small
\caption{Dataset statistics over ten categories. The number of hyperedges is reported in the format of: \#nodes / \#hyperedges.}
\label{tab:dataset}
\centering
\begin{tabular}{l|ccc|ccc|cccc}
\hline
\multirow{2}{*}{Category} & \multicolumn{3}{c|}{Number of Nodes} & \multicolumn{3}{c|}{Number of Edges}                    & \multicolumn{4}{c}{Number of Hyperedges}                        \\ \cline{2-11} 
                          & \#C       & \#P         & \#A        & \#CP    & \#PA    & Density & P-P\textsubscript{also view} & P-P\textsubscript{also buy} & P-A           & C-P-A      \\ \hline
Arts                      & 980       & 11,625      & 2,184      & 50,652  & 28,932  & 7.2$\times 10^{-4}$                              & 970/624          & 1,448/1,248     & 13,809/11,643 & 14,789/979 \\
Books                     & 410       & 16,220      & 255        & 48.271  & 23,438  & 5.03$\times 10^{-4}$                                 & 1,247/1,433      & 2,432/2,550     & 16,475/16,222 & 16,885/409 \\
Cellphones                & 145       & 8,499       & 1,484      & 27,620  & 20,329  & 9.35$\times 10^{-4}$                               & 366/362          & 171/160         & 9,983/8,507   & 10,128/144 \\
Giftcards                 & 5         & 131         & 11         & 378     & 311     & 0.06                                  & 17/20            & 19/32           & 142/130       & 147/1      \\
Grocery                   & 742       & 18,315      & 4,686      & 75,362  & 47,745  & 4.37$\times 10^{-4}$                                & 3,162/2,431      & 3,392/3,314     & 23,001/4,686  & 23,743/741 \\
Industrial                & 433       & 3002        & 1573       & 12,429  & 8,453   & 1.67$\times 10^{-3}$                                & 152/106          & 210/205         & 4,539/3,063   & 5,008/432  \\
Pet                       & 508       & 14,299      & 2,575      & 64,947  & 46,370  & 7.34$\times 10^{-4}$                                & 1,614/1,670      & 820/600         & 16,874/14,675 & 17,382/507 \\
Software                  & 303       & 254         & 98         & 1,182   & 607     & 8.35$\times 10^{-3}$                                & 19/20            & 2/1             & 352/287       & 655/302    \\
Tools                     & 975       & 34,076      & 7,538      & 143,683 & 101,475 & 2.7$\times 10^{-4}$                                & 3,176/2,648      & 1,998/1,704     & 41,614/34,705 & 42,589/974 \\
Videogames                & 910       & 731         & 353        & 4,446   & 2,152   & 3.32 $\times 10^{-3}$                               & 113/139          & 14/9            & 1,084/752     & 1,994/909  \\ \hline
\end{tabular}
\end{table*}

To support representation learning on the constructed heterogeneous hypergraphs in the previous step, we design a heterogeneous hypergraph relation learning module (shown in Figure~\ref{fig:framework}(b) in HyperPAVE to explore the complex higher-order relationships based on many-to-many node message passing in the product graph by taking full advantage of the structure information in Figure~\ref{fig:framework}(a).
%Apart from conventional graph-based neural networks models,
HyperPAVE learns node representations with two different aggregation functions:
\begin{align}
h_{v_i}^{l}&=AGGR_{edge}^{l}\begin{pmatrix}
h_{v_i}^{l-1},\left \{ h_{e_j}^{l}|\forall _{e_{j}}\in \mathcal{E}_{i} \right \}
\end{pmatrix} \\
h_{e_j}^{l} &= AGGR_{node}^{l}\begin{pmatrix}
\left \{ h_{v_k}^{l-1}|\forall _{v_{k}}\in e_{j} \right \}
\end{pmatrix}
\end{align}
where $AGGR$ is the aggregation function, $\mathcal{E}_{i}$ is the hyperedge sets connected to node $v_i$ and $h_{e_j}^{l}$ is the representation of hyperedge $e_{j}$ in layer $l$. 
Since not all the nodes in a hyperedge will contribute equally, the message passing is calculated from nodes to hyperedges:
\begin{align}
  \alpha _{v_i}^{e_i} &= \frac{exp(LeakyReLU(w_{1}^{T}\cdot h_{v_i}^{l-1}))}{\sum_{v \in V_{e_i}}^{}exp(LeakyReLU(w_{1}^{T}\cdot h_{v}^{l-1}))}   \\
  h_{e_i}^{l} &= ||_{n=1}^{N}\sigma (\sum_{v \in V_{e_i}}^{}\alpha_{v}^{e_i}\cdot h_{v}^{l-1})
\end{align}
where $\alpha _{v_i}^{e_i}$ is the weight factor of node $v_i$ to hyperedge $e_i$, $V_{e_i}$ is the node set of hyperedges $e_i$, $w_{1}^{T}$ is a trainable attention parameter, $||$ denotes concatenation with $N$ heads, and $\sigma$ is a non-linear function. $h_{e_i}^{l}$ is the $l^{th}$ layer of hyperedge representation. Similarly, the message passing from hyperedges to nodes is calculated as:
\begin{align}
    \alpha _{e_i}^{v_i} &= \frac{exp(LeakyReLU(w_{2}^{T}\cdot (h_{v_i}^{l-1} || h_{e_i}^{l-1})))}{\sum_{e \in \mathcal{E}_{v_i}}^{}exp(LeakyReLU(w_{1}^{T}\cdot (h_{v_i}^{l-1} || h_{e}^{l-1})))} \\
    h_{v_i}^{l} &= ||_{n=1}^{N}\sigma (\sum_{e \in \mathcal{E}_{v_i}}^{}\alpha_{e}^{v_i}\cdot h_{e}^{l-1})
\end{align}
where $\alpha _{e_i}^{v_i}$ is the weight factor of hyperedge $e_i$ to node $v_i$, $\mathcal{E}_{v_{i}}$ is the connected hyperedge set of node $v_i$. $w_{2}^T$ is a trainable attention parameter and $h_{v_i}^{l}$ is the $l^{th}$ layer of node representation, which includes the information from the hyperedge $\mathcal{E}$.

%We first divide HyperPAVE into different channels $C=\left \{ C_{1}, C_{2}, \cdots, C_{n} \right \}$ based on the different types of hyperedges introduced in Sec.~\ref{sec:construction}.
%For each channel $C_{i}$, 
\begin{comment}
%Hypergraph Convolution
Hypergraph convolution operator~\cite{bai2021hypergraph} can be defined as:
\begin{equation}
    X^{(l+1)}=\sigma (D^{-1/2}_{v}HWD_{e}^{-1}H^{T}D^{-1/2}_{v}X^{(l)}\Omega^{(l)})
\end{equation}
where $X^{(l)} \in \mathbb{R}^{\mathcal{E} \times F^{(l)}}$ and $X^{(l+1)} \in \mathbb{R}^{\mathcal{E} \times F^{(l+1)}}$ are the embedding of vertex feature of the $(l)^{th}$ and $(l+1)^{th}$ layer, respectively.
$\sigma(\cdot)$ is a nonlinear activation function.
$D_{v}$ and $D_{e}$ are the degree matrix of the nodes and hyperedges in a hypergraph, respectively.
$H\in \mathbb{R}^{\left | \mathcal{V} \right | \times \left | \mathcal{E} \right | }$ is the incidence matrix.
$\Omega \in \mathbb{R}^{F^{(l)} \times F^{(l+1)}}$ is the trainable weight matrix between the $(l)^{th}$ and $(l+1)^{th}$ layer, where $F^{(l)}$ and $F^{(l+1)}$ are the dimensions of the $(l)^{th}$ and $(l+1)^{th}$ hidden layer, respectively. 

At each iteration, nodes first aggregate information from their neighbors within a specific hyperedge.
This is repeated over all the hyperedges across all the $L$ layers. The trainable weight matrices $W_l$ with $l \in L$ are used to aggregate information across the feature dimension and propagate it through the hypergraph. 
\end{comment}

\paragraph{Fusion Module}
Instead of directly adding a readout layer and a linear prediction layer after obtaining the $L$ layers node representations~\cite{xu2023correlative}, we argue that different types of hyperedges from $\mathcal{T}_{e}$ have different importance to the final node representations. Thus, we propose fusion modules to fuse node representations learned from different hypergraphs constructed in Sec.~\ref{sec:construction}. The updated node representations for product node $\hat{h_{v_{p_i}}}$ and aspect node $\hat{h_{v_{a_i}}}$ are:
\begin{equation}
    \hat{h_{v_{p_i}}} = \alpha \cdot h_{v_{p_i}}^{\mathcal{E^P}} + \beta \cdot h_{v_{p_i}}^{\mathcal{E^C}} + (1-\alpha-\beta)(\gamma \cdot h_{v_{p_i}}^{\mathcal{E^V}} + (1-\gamma) \cdot h_{v_{p_i}}^{\mathcal{E^B}})
\end{equation}
\begin{equation}
    \hat{h_{v_{a_i}}} = \delta \cdot h_{v_{a_i}}^{\mathcal{E^P}} + (1 - \delta) \cdot h_{v_{a_i}}^{\mathcal{E^C}}
\end{equation}
where $h_{v_{p_i}}^{\mathcal{E^P}}$, $\cdot h_{v_{p_i}}^{\mathcal{E^C}}$, $h_{v_{p_i}}^{\mathcal{E^V}}$, $h_{v_{p_i}}^{\mathcal{E^B}}$ are product node representations and $h_{v_{a_i}}^{\mathcal{E^P}}$, $h_{v_{a_i}}^{\mathcal{E^C}}$ are aspect node representations from different hyperedges in Equ.~\ref{equ:hyperedge}, respectively. $\alpha$, $\beta$, $\gamma$, and $\delta$ are weights learnt from the validation sets. They are different for different categories of the dataset. These weights are also explored and studied in Sec.~\ref{sec:sensitivity}.
After the above fusion steps, the node embeddings contain the features from neighbors defined by different hyperedges $\mathcal{T}_{e}$, which can well capture the high-order relations communicated among different types of nodes and hyperedges.

\subsubsection{Inductive Link Prediction}
After heterogeneous hypergraph relation learning, each node includes the higher-order features related to user behavior and product inventory information.
Then, all the nodes go through $L$ GNN layers to compute the final node representations.
After generating the final embeddings of $\tilde{h_{v_p}}$ and $\tilde{h_{v_a}}$, the likelihood of the link between product $p$ and aspect $a$ is measured by the cosine similarity to decide the possibility $\hat{R}_{ij}$ of whether product $p_i$ will have the aspect $a_j$: %
\begin{equation}
    f_{score}((\tilde{h_{v_p}})_i, (\tilde{h_{v_a}})_j) = \frac{(\tilde{h_{v_p}})_i \cdot (\tilde{h_{v_a}})_j}{\left \| (\tilde{h_{v_p}})_i \right \| \left \| (\tilde{h_{v_a}})_j \right \|}
\end{equation}
%where $Sim()$ is a similarity measure function, which is cosine similarity in our model.

We use the negative sampling strategy introduced in Sec.~\ref{sec:zero_sampling} to train HyperPAVE and employ a binary cross entropy loss to optimize our model:
\begin{equation}
    \mathcal{L} = \sum_{p_{i} \in P, a_{i} \in A}^{}R_{ij}log\hat{R}_{ij}+(1-R_{ij})(1-log\hat{R}_{ij})
\end{equation}

\begin{table*}[]
\small
\caption{Experimental Results F1 / mAP (\%) of multi-label zero-shot learning over ten categories on MAVE. The results are reported as mean $\pm$ standard deviation over ten times of experiments. The best results are in bold.}
\label{tab:main}
\centering
\tabcolsep=0.11cm
\begin{tabular}{lccccc}
\hline
                   & Arts                             & Books                           & Cellphones                      & Giftcards                       & Grocery                          \\ \hline
BERT-MLC~\cite{chen-etal-2022-extreme}           & 24.11$\pm$0.09 / 10.31$\pm$0.16  & 36.72$\pm$0.08 / 27.17$\pm$0.37 & 22.92$\pm$0.25 / 28.67$\pm$0.37 & 36.54$\pm$0.07 / 41.15$\pm$0.08 & 19.74$\pm$0.24 / 12.07$\pm$0.09  \\
Bart~\cite{lewis2019bart}              & 27.88$\pm$0.36 /  23.16$\pm$0.46 & 38.82$\pm$0.44 / 44.90$\pm$0.20 & 32.71$\pm$0.35 / 24.54$\pm$0.37 & 15.73$\pm$0.19 / 8.75$\pm$0.36  & 10.80$\pm$0.18 / 6.95$\pm$0.12   \\
T5\textsubscript{small}~\cite{raffel2020exploring}                 & 30.85$\pm$0.31 / 23.16$\pm$0.17  & 36.17$\pm$0.45 / 42.60$\pm$0.13 & 30.95$\pm$0.31 / 24.27$\pm$0.30 & 10.14$\pm$0.26 / 8.08$\pm$0.23                           & 23.53$\pm$0.27 / 17.32$\pm$0.25  \\
HGCN~\cite{ragesh2021hetegcn}               & 16.87$\pm$0.10 / 25.30$\pm$0.33   & 39.39$\pm$0.18 / 37.40$\pm$0.12 & 17.23$\pm$.24 / 14.67$\pm$0.31  & 30.92$\pm$0.07 / 45.42$\pm$0.06 & 25.60$\pm$0.13 / 39.77$\pm$ 0.18 \\
HAN~\cite{wang2019heterogeneous}                & 14.26$\pm$0.10 / 26.42$\pm$0.25  & 43.73$\pm$0.16 / 49.48$\pm$0.16 & 22.49$\pm$0.20 / 33.69$\pm$0.17 & 42.47$\pm$0.31 / 54.05$\pm$0.13 & 17.23$\pm$0.20 / 34.67$\pm$0.24  \\
HGT~\cite{hu2020heterogeneous}                & 30.81$\pm$0.13 / 38.53$\pm$0.16  & 48.06$\pm$0.11 / 41.67$\pm$0.17 & 14.53$\pm$0.16 / 23.73$\pm$0.20 & 42.30$\pm$0.40 / 42.39$\pm$0.19 & 27.30$\pm$0.11 / 40.76$\pm$0.24  \\
HGNN+~\cite{9795251}               & 27.90$\pm$0.28 / 36.91$\pm$0.13  & 46.79$\pm$0.20 / \textbf{58.33$\pm$0.15} & 32.10$\pm$0.17 / \textbf{36.40$\pm$0.26} & 37.18$\pm$0.07 / 57.20$\pm$0.04 & 32.40$\pm$0.14 / 38.60$\pm$0.15  \\
HyperGCN~\cite{yadati2019hypergcn}           & 20.20$\pm$0.17 / 38.45$\pm$0.21  & 48.97$\pm$0.13 / 45.18$\pm$0.16 & 20.90$\pm$0.25 / 26.00$\pm$0.40 & \textbf{52.74$\pm$0.19} / 45.97$\pm$0.09 & \textbf{35.90$\pm$0.22} / 42.20$\pm$0.21  \\
\textbf{HyperPAVE} & \textbf{43.33$\pm$0.22 / 40.99$\pm$0.18}  & \textbf{49.75$\pm$0.18} / 56.45$\pm$0.11 & \textbf{39.01$\pm$0.16} / 35.81$\pm$0.18 & 52.34$\pm$0.22 / \textbf{65.03$\pm$0.13} & 33.43$\pm$0.28 / \textbf{42.71$\pm$0.30}  \\ \hline
                   & Industrial                       & Pet                             & Software                        & Tools                           & Videogames                       \\ \hline
BERT-MLC~\cite{chen-etal-2022-extreme}           & 10.94$\pm$0.19 / 6.69$\pm$0.16   & 18.14$\pm$0.55 / 12.08$\pm$0.16 & 27.76$\pm$0.09 / 25.37$\pm$0.09 & 20.43$\pm$0.26 / 18.41$\pm$0.17 & 11.86$\pm$0.31 / 9.66$\pm$0.35   \\
BART~\cite{lewis2019bart}              & 10.78$\pm$0.32 / 7.84$\pm$0.32   & 12.50$\pm$0.25 / 10.42$\pm$0.67 & 22.50$\pm$0.03 / 20.00$\pm$0.02 & 11.11$\pm$0.16 / 6.25$\pm$0.09  & 23.57$\pm$0.32 / 20.02$\pm$0.25  \\
T5\textsubscript{small}~\cite{raffel2020exploring}                 & 15.81$\pm$0.47 / 15.35$\pm$0.16  & 25.28$\pm$0.20 / 25.72$\pm$0.26 & 26.19$\pm$0.42 / 24.60$\pm$0.31 & \textbf{37.78$\pm$0.26} / 22.46$\pm$0.52 & 14.41$\pm$0.15 / 9.90$\pm$0.27   \\
HGCN~\cite{ragesh2021hetegcn}               & 10.67$\pm$0.24 / 14.60$\pm$0.14  & 17.62$\pm$0.15 / 24.63$\pm$0.24 & 19.29$\pm$0.22 / 30.97$\pm$0.15 & 18.07$\pm$0.20 / 39.32$\pm$0.18 & 8.78$\pm$0.40 / 13.61$\pm$0.25   \\
HAN~\cite{wang2019heterogeneous}                & 15.35$\pm$0.20 / 30.45$\pm$0.50  & 16.82$\pm$0.13 / 23.33$\pm$0.25 & 28.24$\pm$0.31 / 29.03$\pm$0.14 & 19.78$\pm$0.03 / 41.40$\pm$0.14 & 9.68$\pm$0.16 / 16.29$\pm$0.21   \\
HGT~\cite{hu2020heterogeneous}                & 21.09$\pm$0.13 / 23.20$\pm$0.16  & 18.02$\pm$0.13 / 23.66$\pm$0.20 & 30.15$\pm$0.20 / 27.16$\pm$0.08 & 13.61$\pm$0.18 / 35.23$\pm$0.22 & 14.75$\pm$0.05 / 19.97$\pm$0.11  \\
HGNN+~\cite{9795251}               & 25.90$\pm$0.26 / 28.60$\pm$0.12  & 27.60$\pm$0.14 / 35.58$\pm$.16  & 39.90$\pm$0.26 / 28.76$\pm$.16  & 31.00$\pm$0.15 / 42.20$\pm$0.23 & 10.35$\pm$0.11 / 17.21$\pm$0.08  \\
HyperGCN~\cite{yadati2019hypergcn}           & \textbf{29.20$\pm$0.13} / 33.20$\pm$.11   & 22.20$\pm$0.12 / 31.37$\pm$0.14 & 42.10$\pm$0.31 / 38.70$\pm$0.13 & 31.10$\pm$0.18 / 44.05$\pm$0.19 & 10.90$\pm$0.13 / 15.30$\pm$0.10  \\
\textbf{HyperPAVE}          & 27.70$\pm$0.10 / \textbf{33.29$\pm$0.17}  & \textbf{28.45$\pm$0.13 / 38.46$\pm$0.20} & \textbf{47.62$\pm$0.21 / 51.64$\pm$0.10} & 34.00$\pm$0.28 / \textbf{47.83$\pm$0.29} & \textbf{25.31$\pm$0.19 / 21.19$\pm$0.17}  \\ \hline
\end{tabular}
\end{table*}

Note that HyperPAVE follows the multi-label zero-shot settings in Sec.~\ref{sec:zero_sampling} to eliminate the mandatory access of testing node features during training, making the model access the inductive inference ability.
For unseen attribute values (aspects) and products, we can directly feed their corresponding contextual node embeddings by fine-tuned BERT encoder to HyperPAVE instead of representing product and aspect nodes with one-hot vectors.
Then, we only conduct message-passing and compute the probability of connections between the product node and the aspect node.
Hence, we can handle the newly added products and attribute values in an inductive way instead of retraining the model.

\section{Experiments}
\subsection{Experimental Setup}
\subsubsection{Dataset}
% Please add the following required packages to your document preamble:
% \usepackage{multirow}
We evaluate our model~\footnote{The code is available: \url{https://github.com/gjiaying/HyperPAVE}.} over ten different categories (Arts, Books, Cellphones, etc) of a public dataset MAVE~\cite{yang2022mave}, which is a large e-Commerce dataset derived from Amazon Review Dataset~\cite{ni2019justifying}.
To simulate the zero-shot situation, we reconstruct the dataset into multi-label zero-shot learning settings followed by Sec.~\ref{sec:zero_sampling}, where there is no overlap of products and attribute values between the training set and validation/testing set.
%The dataset statistics is shown in Table~\ref{tab:dataset}. 
Note that each time we train the model, the dataset will be randomly re-splitted for the zero-shot setting, so we report the whole data statistics in Table~\ref{tab:dataset}. 
A sample of data statistics for training, validation, and testing sets for each category is shown in Appendix~\ref{sec:appendix_dataset}.

\subsubsection{Evaluation Metrics}
%Single-label ZSL usually evaluates performance by accuracy.
Following other AVE tasks in the multi-label zero-shot setting~\cite{shinzato2023unified}, we choose to report macro-F1 and mAP (mean Average Precision) compared with classification and generation-based models in the main results as F1 score is the balance of both precision and recall.
%Generative and classification models
In Sec.~\ref{sec:ablation} ablation study, we also report AUC (Area Under Curve), Hits@K, NDCG@K (Normalized Discounted Cumulative Gain), and MRR (Mean Reciprocal Ran), which are widely used metrics in graph-based recommendation tasks~\cite{fan2023zero, han2023intra, li2022heterogeneous}.
We also report training time to evaluate the efficiency in Sec.~\ref{sec:efficiency} efficiency study.

\subsubsection{Baselines}
We compare our proposed model HyeprPAVE with the following baselines in the zero-shot setting:
\begin{itemize}
    \item Classification-based Models: Original classification-based models do not have any zero-shot abilities. We follow the baseline \textbf{BERT-MLC} in~\cite{shinzato2023unified}, then we add synthetic data for unseen classes (attribute values) following~\cite{chia-etal-2022-relationprompt}. In this way, the zero-shot learning problem is translated into a supervised learning problem.
    \item Generation-based Models: Following generative models in zero-shot AVE task~\cite{shinzato2023unified}, we implement and fine-tune two text-to-text transformer-based encoder decoder architecture models: \textbf{BART}~\cite{lewis2019bart} and \textbf{T5\textsubscript{small}}~\cite{raffel2020exploring}, to generate unseen attribute values directly. 
    \item Graph-based Models~\footnote{Implemented on DHG: https://deephypergraph.com/}: As inductive graph can predict unseen nodes (zero-shot learning), 
    we compare HyperPAVE with three heterogeneous graph neural networks: \textbf{HGCN}~\cite{ragesh2021hetegcn}, \textbf{HAN}~\cite{wang2019heterogeneous}, 
    \textbf{HGT}~\cite{hu2020heterogeneous}, and two representative hypergraph networks:
    \textbf{hyperGCN}~\cite{yadati2019hypergcn}, \textbf{HGNN+}~\cite{9795251}.

\end{itemize}

\subsection{Results and Discussions}
\begin{table*}[]
\small
\caption{Ablation study over HyperPAVE components in the zero-shot setting across three categories on MAVE dataset.}
\label{tab:ablation}
\centering
\begin{tabular}{l|cccccccc}
\hline
                      & F1                        & mAP                       & AUC                       & MRR                       & NDCG                      & Hits@5                     & Hits@10                    & Hits@100                   \\ \hline
                      & \multicolumn{8}{c}{Books}                                                                                                                                                                                                        \\ \hline
nodeID                & 11.54 $\pm$ 1.59          & 28.52 $\pm$ 1.29          & 95.31 $\pm$ 1.15          & 6.64 $\pm$ 1.07           & 48.41 $\pm$ 0.97          & 35.26 $\pm$ 0.63           & 53.85 $\pm$ 0.62           & 99.42 $\pm$ 0.05           \\
BERT                  & 23.87 $\pm$ 1.29          & 38.63 $\pm$ 0.57          & 97.07 $\pm$ 0.60          & 11.39 $\pm$ 0.83          & 57.31 $\pm$ 0.65          & 47.05 $\pm$ 0.42           & 63.59 $\pm$ 0.40           & 100.00 $\pm$ 0.00          \\
BERT (Fine-tuned)     & 28.28 $\pm$ 0.81          & 40.32 $\pm$ 0.59          & 97.87 $\pm$ 0.21          & 14.44 $\pm$ 0.40          & 58.89 $\pm$ 0.41          & 50.90 $\pm$ 0.82           & 78.33 $\pm$ 0.38           & 100.00$\pm$ 0.00           \\
Hyper (Product)  & 30.44 $\pm$ 0.25          & 40.65 $\pm$ 0.41          & 98.03 $\pm$ 0.19          & 14.23 $\pm$ 0.19          & 59.49 $\pm$ 0.30          & 49.36 $\pm$ 0.14           & 80.51 $\pm$ 0.17           & 100.00$\pm$ 0.00           \\
Hyper (Behavior) & 34.46 $\pm$ 0.29          & 35.93 $\pm$ 0.49          & \textbf{98.40 $\pm$ 0.20} & 19.37 $\pm$ 0.27          & 54.23 $\pm$ 0.31          & 63.67 $\pm$ 0.40           & 93.67 $\pm$ 0.24           & 100.00$\pm$ 0.00           \\
\textbf{HyperPAVE}    & \textbf{49.75 $\pm$ 0.18} & \textbf{56.45 $\pm$ 0.11} & 96.47 $\pm$ 0.02          & \textbf{32.99 $\pm$ 0.14} & \textbf{69.35 $\pm$ 0.18} & \textbf{85.27 $\pm$ 0.12}  & \textbf{94.04 $\pm$ 0.08}  & \textbf{100.00 $\pm$ 0.00} \\ \hline
                      & \multicolumn{8}{c}{Giftcards}                                                                                                                                                                                                    \\ \hline
nodeID                & 6.67 $\pm$ 0.21           & 22.35 $\pm$ 0.20          & 41.94 $\pm$ 0.29          & 18.18 $\pm$ 0.30          & 42.92 $\pm$ 0.15          & 25.00 $\pm$ 0.00           & 97.50 $\pm$ 0.08           & 100.00$\pm$ 0.00           \\
BERT                  & 26.41 $\pm$ 0.17          & 44.79 $\pm$ 0.14          & 71.94 $\pm$ 0.05          & 24.15 $\pm$ 0.18          & 62.47 $\pm$ 0.13          & 75.00 $\pm$ 0.00           & 100.00$\pm$ 0.00           & 100.00$\pm$ 0.00           \\
BERT (Fine-tuned)     & 34.43 $\pm$ 0.17          & 41.67 $\pm$ 0.15          & 71.53 $\pm$ 0.16          & 23.17 $\pm$ 0.30          & 59.57 $\pm$ 0.11          & 67.50 $\pm$ 0.12           & 100.00$\pm$ 0.00           & 100.00$\pm$ 0.00           \\
Hyper (Product)  & 39.77 $\pm$ 0.12          & 45.74 $\pm$ 0.10 & 84.55 $\pm$ 0.10          & 35.65 $\pm$ 0.17          & 61.83 $\pm$ 0.08          & 100.00$\pm$ 0.00           & 100.00$\pm$ 0.00           & 100.00$\pm$ 0.00           \\
Hyper (Behavior) & 45.43$\pm$ 0.15          & 60.50 $\pm$ 0.13          & 77.92 $\pm$ 0.12          & 29.13 $\pm$ 0.15          & 73.02 $\pm$ 0.07 & 71.00 $\pm$ 0.12           & 100.00$\pm$ 0.00           & 100.00$\pm$ 0.00           \\
\textbf{HyperPAVE}             & \textbf{52.34 $\pm$ 0.22} & \textbf{65.03 $\pm$ 0.13  }        & \textbf{90.08 $\pm$ 0.05} & \textbf{44.56 $\pm$ 0.16} & \textbf{75.07 $\pm$ 0.11 }         & \textbf{100.00 $\pm$ 0.00} & \textbf{100.00 $\pm$ 0.00} & \textbf{100.00 $\pm$ 0.00} \\ \hline
                      & \multicolumn{8}{c}{Pets}                                                                                                                                                                                                         \\ \hline
nodeID                & 6.95 $\pm$ 0.82           & 13.46 $\pm$ 0.92          & 98.51 $\pm$ 0.27          & 7.47 $\pm$ 0.19           & 42.17 $\pm$ 0.62          & 30.33 $\pm$ 0.46           & 50.00 $\pm$ 0.13           & 96.15 $\pm$ 0.10           \\
BERT                  & 9.93 $\pm$ 0.30           & 19.93 $\pm$ 0.51          & 99.73 $\pm$ 0.20          & 6.71 $\pm$ 0.27           & 41.16 $\pm$ 0.68          & 31.67$\pm$ 0.54            & 65.00 $\pm$ 0.50           & 100.00$\pm$ 0.00           \\
BERT (Fine-tuned)     & 12.12 $\pm$ 0.29          & 19.99 $\pm$ 0.74          & 99.39 $\pm$ 0.11          & 10.79 $\pm$ 0.37          & 40.52 $\pm$ 0.61          & 25.00$\pm$ 0.29            & 56.67 $\pm$ 0.14           & 100.00$\pm$ 0.00           \\
Hyper (Product)  & 17.58 $\pm$ 0.26          & 37.40 $\pm$ 0.32          & 99.03 $\pm$ 0.09          & 16.45 $\pm$ 0.19          & 45.67 $\pm$ 0.21          & 41.67 $\pm$ 0.17           & \textbf{71.67 $\pm$ 0.15}  & 98.33 $\pm$ 0.12           \\
Hyper (Behavior) & 18.66 $\pm$ 0.14          & 24.71 $\pm$ 0.14          & 99.16 $\pm$ 0.10          & 19.01 $\pm$ 0.28          & 42.38 $\pm$ 0.35          & 36.07 $\pm$ 0.22           & 65.00 $\pm$ 0.09           & 100.00$\pm$ 0.00           \\
\textbf{HyperPAVE}    & \textbf{28.45 $\pm$ 0.13} & \textbf{38.46 $\pm$ 0.20} & \textbf{99.82 $\pm$ 0.06} & \textbf{29.92 $\pm$ 0.13} & \textbf{61.55 $\pm$ 0.20} & \textbf{56.67 $\pm$ 0.09}  & 67.77 $\pm$ 0.03           & \textbf{100.00$\pm$ 0.00}  \\ \hline
\end{tabular}
\end{table*}

\subsubsection{Main Results}
%The experiment results of multi-label zero-shot learning across ten different categories on the MAVE dataset are shown in Table~\ref{tab:main}.
From the results shown in Table~\ref{tab:main} and data statistics shown in Table~\ref{tab:dataset}, we observe that:

(1) The classification-based model generally performs worst among all models. BERT-MLC, which uses synthetic data for zero-shot prediction, only has competitive performance to generation-based models when the class number (\#A) is small. We conjecture that as the number of classes grows, BERT-MLC needs to make distinctions among more classes, making it harder to find clearer decision boundaries. 
The average micro F1 of BART and T5~\textsubscript{small} across all ten categories is worse than T5~\textsubscript{base} in~\cite{shinzato2023unified}. This is because T5~\textsubscript{base} is pre-trained over 220 million parameters whereas T5~\textsubscript{small} has only 60 million parameters.
Generation-based models perform much better than classification-based models in most cases. BART and T5~\textsubscript{small} show different performances over different categories. They can achieve similar performance with HyperPAVE when the dataset size is large enough. 
(2) Combining inductive graph-based models with LLM encoders can perform zero-shot prediction and achieve competitive performance with generative models. This inspires us that instead of fine-tuning the popular generative models~\cite{shinzato2023unified, zhang2023pay, roy2022exploring, roy2021attribute} to extract attribute values, the inductive graph for link prediction can also be explored for zero-shot prediction. 
Besides, using the attention mechanism shows better performance than using fixed and uniform weights for aggregation. This is probably because assigning different weights to neighboring nodes can capture varying levels of influence.
(3) Compared with graph-based baselines, adding complex structured data to capture higher-order relationships demonstrates significant performance improvement over all ten categories.
This is probably because hyperedges can model relationships that go beyond pairwise connections, resulting in more semantic node representations.
Besides, our proposed model HyperPAVE achieves the best performance among all models in most categories, indicating that our proposed hypergraph construction from both user behavior data and product inventory data is important and worth recording and exploring. The effectiveness of different hyperedges is studied in Sec.~\ref{sec:ablation}.

\subsubsection{Ablation Study}~\label{sec:ablation}
To evaluate the performance of each component in HyperPAVE, we conduct an ablation study over three categories in the zero-shot setting.
%Based on the average number of aspects per product shown in the last column of Table~\ref{tab:datasplits}, books, giftcards, and pet categories have the smallest, medium, and largest number of aspects for each product, respectively.
%Thus, we chose these categories to report ablation studies due to the limited space.
Table~\ref{tab:ablation} shows the performance of each component in HyperPAVE. More results are shown in Table~\ref{tab:ablations} in the Appendix.
We have the following observations based on Table~\ref{tab:ablation}:
(1) Adding node features can significantly improve the performance. We perform a model `nodeID', which doesn't use any pre-trained encoder for providing node features.
The model `nodeID' uses a simple embedding-lookup encoder, mapping each node to a unique low-dimensional vector. We can observe that among all models, `nodeID' shows the worst performance. After adding node features, such as BERT or fine-tuned BERT, the performance increases significantly. We think that this is because, for link prediction in the zero-shot setting, pre-trained embeddings provide richer and more semantically meaningful representations for node features in graphs than simple one-hot encoding.
(2) Fine-tuning the pre-trained encoders for node features results in a big performance improvement when the dataset (graph) is large enough.
This is reasonable because a larger dataset (graph with more nodes) provides more diverse and representative data, enabling better generalization for unseen nodes in the zero-shot setting.
However, as shown in Sec.~\ref{sec:efficiency}, fine-tuning the pre-trained encoder may result in more time for model training.
A balance of model performance and efficiency needs to be considered for different tasks/situations.
(3) We explore the importance of different hypergraphs. We find out that adding different hyperedges built from user behavior data or product inventory data results in a significant performance improvement. 
We conjecture that this is because different hyperedges capture more complex higher-order information than the original binary-relation graph.
For example, hyperedge 'P-P\textsubscript{also\_view}' built from user behavior data includes information on products with potential similar attributes because users may probably view similar products at the same time for their needs.
Hyperedge `C-P-A', built from product inventory data, aggregates all products and aspects in the same sub-category. Attribute values such as `Chew Type: Bones' may only happen in a sub-category of `Dog Treats' instead of `Cat Food'. By using hyperedges, more complex relations can be included in the representation for each single node.

\subsubsection{Efficiency Study}~\label{sec:efficiency}
\begin{table}[]
\small
\caption{Comparison of computational efficiency. The batch size is set to 4.}
\label{tab:computational}
\centering
\begin{tabular}{lcc}
\hline
Model                & Memory Consumption & Model Parameters \\ \hline
Classification-based & 5037MB                 & 110M               \\
Generation-based     & 8305MB / 5831MB        & 140M / 60M         \\
Graph-based (ours)         & 1405MB / 1915 MB       & 5M / 115M          \\ \hline
\end{tabular}
\end{table}
Table~\ref{tab:computational} presents the GPU computational cost and model parameter comparison between classification-based (BERT-MLC), generation-based (BART/T5\textsubscript{small}) and graph-based (nodeID/HyperPAVE) models on Arts category of MAVE.
Different categories (different sizes of graphs) may result in a slight difference.
From the reported results, we can find that compared with classification or generation-based models, our proposed graph-based model HyperPAVE, has a significant computational advantage in terms of memory consumption.
The main reason is that the zero-shot ability of generative LLMs is based on their extensive pretraining and understanding of the diverse data. 
When fine-tuning these LLMs, large quantities of model parameters need to be updated, resulting in a huge GPU memory consumption cost.
However, the zero-shot ability of HyperPAVE results from the inductive inference that can generalize to unseen product and aspect nodes without retraining the whole model.
The inductive HyperPAVE divides the hypergraphs into batches and only consumes per-batch memory when training.
Note that for the classification model BERT-MLC, preprocessing steps for generating synthetic data from generation models are required to predict unseen aspects. We have not counted the computational cost for these preprocessing steps.

\begin{figure}[htp] 
 \center{\includegraphics[height=6cm,width=8cm]{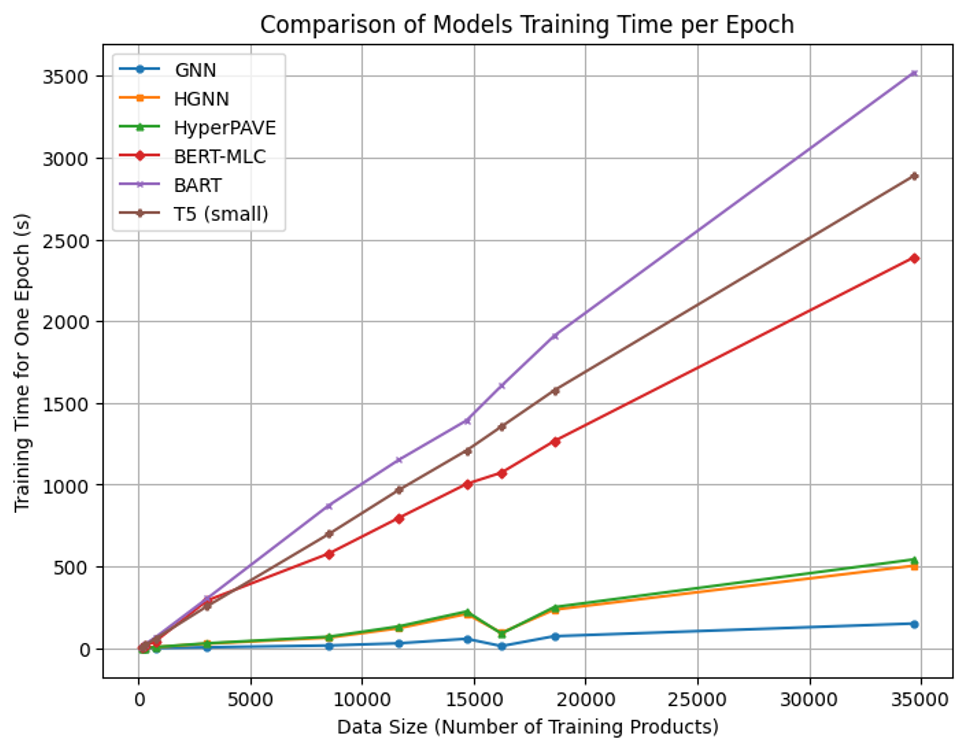}}
 \caption{\label{fig:time} Time Efficiency Performance (GPU Time of Model Learning in Seconds for One Training Epoch).}
 \end{figure}

To evaluate the computation time of our graph-based model HyperPAVE and other classification-based and generation-based models, we record the model training time for one epoch in seconds across the ten categories on MAVE as shown in Figure~\ref{fig:time}. 
All models use the same input max\_length and batch size for training.
From Figure~\ref{fig:time}, we observe that graph-based models show better model training efficiency.
Compared with other graph-based models (i.e. GNN, HGNN), HyperPAVE can achieve the best prediction performance as shown in Table~\ref{tab:main} with only sacrificing a little more time for training as shown in Figure~\ref{fig:time}.
%but other generation or classification-based models need much more time for training because more model parameters are needed to update.
The slopes of BART, T5, and BERT-MLE are much larger than graph-based models, indicating that much more time is needed for training or fine-tuning with the increase in dataset size when updating the model parameters.
More details are shown in Table~\ref{tab:time} in Appendix~\ref{sec:more_experiments}.

\subsubsection{Parameter Sensitivity Analysis}~\label{sec:sensitivity}
\begin{figure}[!htbp]
\centering
{
\includegraphics[width=4.0cm]{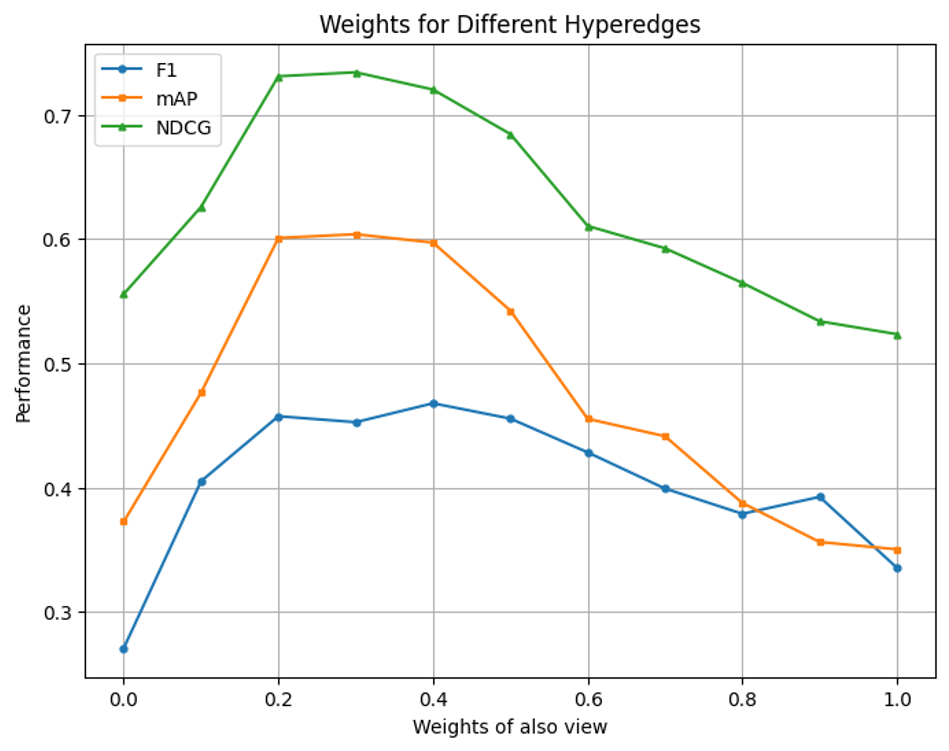}
%\caption{fig1}
}
\hspace{-2mm}
{
\includegraphics[width=4.0cm]{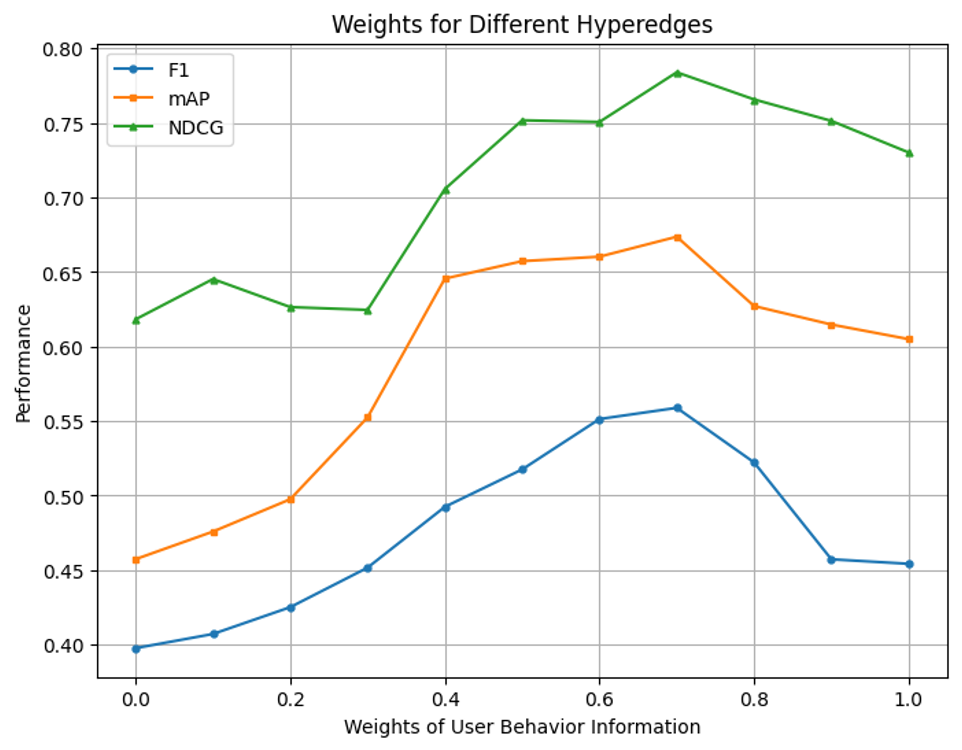}
}
\caption{Effects on weights of different hyperedges on the category of giftcards.}
\label{fig:parameter} 
\end{figure}

The key hyperparameters of HyperPAVE are the weights of the different hyperedges. 
%In other words, we explore 
Thus, we explore the importance of different types of hyperedges in the category of Giftcards as shown in Figure~\ref{fig:parameter}.

The left figure explores the weights of `P-P\textsubscript{also view}' and `P-P\textsubscript{also buy}' hyperedges from user behavior information.
The right figure explores the weights of user behavior hyperedges (P-P) and product inventory hyperedges (`P-A' and `C-P-A'). 
From Figure~\ref{fig:parameter}, we observe that both `P-P\textsubscript{also view}' and `P-P\textsubscript{also buy}' contribute to the model's performance. The best weight for `P-P\textsubscript{also view}' falls in the [0.2, 0.5] interval, which means `P-P\textsubscript{also buy}' is slightly more important than `P-P\textsubscript{also view}'. This is probably because `P-P\textsubscript{also buy}' records users' history preference while `P-P\textsubscript{also view}' may include some noise such as accidental clicks. 
We can also observe from the right ~\ref{fig:parameter} that the best weight for user behavior data falls in the [0.6, 0.8] interval, indicating that user behavior is much more important than product inventory data. As shown in Table~\ref{tab:dataset}, the number of user behavior hyperedges is much smaller than the number of product inventory hyperedges (`P-A' and `C-P-A'). %However, user behavior information contributes more to the
But they show more importance in Figure~\ref{fig:parameter}, demonstrating that user behavior information is worth recording and exploring for extracting unseen attribute values for new products.

\section{Conclusion and Future Work}
In this paper, we formulate the AVE task in a zero-shot learning scenario to identify unseen attribute values from new products with no corresponding labeled data available for training.
We propose an inductive heterogeneous hypergraph (HyperPAVE) for multi-label zero-shot attribute value extraction. 
Specifically, the heterogeneous hypergraph captures the higher-order relationships among users and products, and the inductive mechanism infers the future connections between unseen nodes.
Extensive experimental results on ten different categories across the public dataset MAVE demonstrate that our proposed model HyperPAVE outperforms other state-of-the-art classification-based and generation-based models.
The ablation study validates the efficiency and effectiveness of different hypergraphs constructed from user behavior and product inventory data.
%Future Work
We plan to explore the following directions in future work:
(1) Including multimodal features (i.e. product images) as node attributes to capture more semantic information from the products.
(2) Building dynamic graphs by including timestamps to make the product graph adapt to the developing market.

%%
%% The next two lines define the bibliography style to be used, and
%% the bibliography file.

\bibliographystyle{ACM-Reference-Format}
\balance
\bibliography{sample-base}

%%% -*-BibTeX-*-
%%% Do NOT edit. File created by BibTeX with style
%%% ACM-Reference-Format-Journals [18-Jan-2012].

\begin{thebibliography}{80}

%%% ====================================================================
%%% NOTE TO THE USER: you can override these defaults by providing
%%% customized versions of any of these macros before the \bibliography
%%% command.  Each of them MUST provide its own final punctuation,
%%% except for \shownote{}, \showDOI{}, and \showURL{}.  The latter two
%%% do not use final punctuation, in order to avoid confusing it with
%%% the Web address.
%%%
%%% To suppress output of a particular field, define its macro to expand
%%% to an empty string, or better, \unskip, like this:
%%%
%%% \newcommand{\showDOI}[1]{\unskip}   % LaTeX syntax
%%%
%%% \def \showDOI #1{\unskip}           % plain TeX syntax
%%%
%%% ====================================================================

\ifx \showCODEN    \undefined \def \showCODEN     #1{\unskip}     \fi
\ifx \showDOI      \undefined \def \showDOI       #1{#1}\fi
\ifx \showISBNx    \undefined \def \showISBNx     #1{\unskip}     \fi
\ifx \showISBNxiii \undefined \def \showISBNxiii  #1{\unskip}     \fi
\ifx \showISSN     \undefined \def \showISSN      #1{\unskip}     \fi
\ifx \showLCCN     \undefined \def \showLCCN      #1{\unskip}     \fi
\ifx \shownote     \undefined \def \shownote      #1{#1}          \fi
\ifx \showarticletitle \undefined \def \showarticletitle #1{#1}   \fi
\ifx \showURL      \undefined \def \showURL       {\relax}        \fi
% The following commands are used for tagged output and should be
% invisible to TeX
\providecommand\bibfield[2]{#2}
\providecommand\bibinfo[2]{#2}
\providecommand\natexlab[1]{#1}
\providecommand\showeprint[2][]{arXiv:#2}

\bibitem[Chen and Li(2021)]%
        {chen-li-2021-zs}
\bibfield{author}{\bibinfo{person}{Chih-Yao Chen} {and}
  \bibinfo{person}{Cheng-Te Li}.} \bibinfo{year}{2021}\natexlab{}.
\newblock \showarticletitle{{ZS}-{BERT}: Towards Zero-Shot Relation Extraction
  with Attribute Representation Learning}. In
  \bibinfo{booktitle}{\emph{Proceedings of the 2021 Conference of the North
  American Chapter of the Association for Computational Linguistics: Human
  Language Technologies}}. \bibinfo{publisher}{Association for Computational
  Linguistics}, \bibinfo{address}{Online}, \bibinfo{pages}{3470--3479}.
\newblock
\urldef\tempurl%
\url{https://doi.org/10.18653/v1/2021.naacl-main.272}
\showDOI{\tempurl}


\bibitem[Chen et~al\mbox{.}(2021)]%
        {ijcai2021p597}
\bibfield{author}{\bibinfo{person}{Jiaoyan Chen}, \bibinfo{person}{Yuxia Geng},
  \bibinfo{person}{Zhuo Chen}, \bibinfo{person}{Ian Horrocks},
  \bibinfo{person}{Jeff Z.~Pan}, {and} \bibinfo{person}{Huajun Chen}.}
  \bibinfo{year}{2021}\natexlab{}.
\newblock \showarticletitle{Knowledge-aware Zero-Shot Learning: Survey and
  Perspective}. In \bibinfo{booktitle}{\emph{Proceedings of the Thirtieth
  International Joint Conference on Artificial Intelligence, {IJCAI-21}}},
  \bibfield{editor}{\bibinfo{person}{Zhi-Hua Zhou}} (Ed.).
  \bibinfo{publisher}{International Joint Conferences on Artificial
  Intelligence Organization}, \bibinfo{pages}{4366--4373}.
\newblock
\urldef\tempurl%
\url{https://doi.org/10.24963/ijcai.2021/597}
\showDOI{\tempurl}
\newblock
\shownote{Survey Track}.


\bibitem[Chen and Zhang(2020)]%
        {chen2020hgmf}
\bibfield{author}{\bibinfo{person}{Jiayi Chen} {and} \bibinfo{person}{Aidong
  Zhang}.} \bibinfo{year}{2020}\natexlab{}.
\newblock \showarticletitle{Hgmf: heterogeneous graph-based fusion for
  multimodal data with incompleteness}. In
  \bibinfo{booktitle}{\emph{Proceedings of the 26th ACM SIGKDD international
  conference on knowledge discovery \& data mining}}.
  \bibinfo{pages}{1295--1305}.
\newblock


\bibitem[Chen et~al\mbox{.}(2023b)]%
        {10.1145/3539597.3570484}
\bibfield{author}{\bibinfo{person}{Mengru Chen}, \bibinfo{person}{Chao Huang},
  \bibinfo{person}{Lianghao Xia}, \bibinfo{person}{Wei Wei},
  \bibinfo{person}{Yong Xu}, {and} \bibinfo{person}{Ronghua Luo}.}
  \bibinfo{year}{2023}\natexlab{b}.
\newblock \showarticletitle{Heterogeneous Graph Contrastive Learning for
  Recommendation}. In \bibinfo{booktitle}{\emph{Proceedings of the Sixteenth
  ACM International Conference on Web Search and Data Mining}} (Singapore,
  Singapore) \emph{(\bibinfo{series}{WSDM '23})}.
  \bibinfo{publisher}{Association for Computing Machinery},
  \bibinfo{address}{New York, NY, USA}, \bibinfo{pages}{544–552}.
\newblock
\showISBNx{9781450394079}
\urldef\tempurl%
\url{https://doi.org/10.1145/3539597.3570484}
\showDOI{\tempurl}


\bibitem[Chen et~al\mbox{.}(2022a)]%
        {chen2022extreme}
\bibfield{author}{\bibinfo{person}{Wei-Te Chen}, \bibinfo{person}{Yandi Xia},
  {and} \bibinfo{person}{Keiji Shinzato}.} \bibinfo{year}{2022}\natexlab{a}.
\newblock \showarticletitle{Extreme Multi-Label Classification with Label
  Masking for Product Attribute Value Extraction}. In
  \bibinfo{booktitle}{\emph{Proceedings of The Fifth Workshop on e-Commerce and
  NLP (ECNLP 5)}}. \bibinfo{pages}{134--140}.
\newblock


\bibitem[Chen et~al\mbox{.}(2022b)]%
        {chen-etal-2022-extreme}
\bibfield{author}{\bibinfo{person}{Wei-Te Chen}, \bibinfo{person}{Yandi Xia},
  {and} \bibinfo{person}{Keiji Shinzato}.} \bibinfo{year}{2022}\natexlab{b}.
\newblock \showarticletitle{Extreme Multi-Label Classification with Label
  Masking for Product Attribute Value Extraction}. In
  \bibinfo{booktitle}{\emph{Proceedings of the Fifth Workshop on e-Commerce and
  NLP (ECNLP 5)}}. \bibinfo{publisher}{Association for Computational
  Linguistics}, \bibinfo{address}{Dublin, Ireland}, \bibinfo{pages}{134--140}.
\newblock
\urldef\tempurl%
\url{https://doi.org/10.18653/v1/2022.ecnlp-1.16}
\showDOI{\tempurl}


\bibitem[Chen et~al\mbox{.}(2023a)]%
        {chen2023integrating}
\bibfield{author}{\bibinfo{person}{Ziyi Chen}, \bibinfo{person}{Yutong Gao},
  \bibinfo{person}{Congyan Lang}, \bibinfo{person}{Lili Wei},
  \bibinfo{person}{Yidong Li}, \bibinfo{person}{Hongzhe Liu}, {and}
  \bibinfo{person}{Fayao Liu}.} \bibinfo{year}{2023}\natexlab{a}.
\newblock \showarticletitle{Integrating Topology beyond Descriptions for
  Zero-shot Learning}.
\newblock \bibinfo{journal}{\emph{Pattern Recognition}} (\bibinfo{year}{2023}),
  \bibinfo{pages}{109738}.
\newblock


\bibitem[Cheng et~al\mbox{.}(2022)]%
        {cheng2022ihgnn}
\bibfield{author}{\bibinfo{person}{Dian Cheng}, \bibinfo{person}{Jiawei Chen},
  \bibinfo{person}{Wenjun Peng}, \bibinfo{person}{Wenqin Ye},
  \bibinfo{person}{Fuyu Lv}, \bibinfo{person}{Tao Zhuang},
  \bibinfo{person}{Xiaoyi Zeng}, {and} \bibinfo{person}{Xiangnan He}.}
  \bibinfo{year}{2022}\natexlab{}.
\newblock \showarticletitle{IHGNN: Interactive Hypergraph Neural Network for
  Personalized Product Search}. In \bibinfo{booktitle}{\emph{Proceedings of the
  ACM Web Conference 2022}}. \bibinfo{pages}{256--265}.
\newblock


\bibitem[Chia et~al\mbox{.}(2022)]%
        {chia-etal-2022-relationprompt}
\bibfield{author}{\bibinfo{person}{Yew~Ken Chia}, \bibinfo{person}{Lidong
  Bing}, \bibinfo{person}{Soujanya Poria}, {and} \bibinfo{person}{Luo Si}.}
  \bibinfo{year}{2022}\natexlab{}.
\newblock \showarticletitle{{R}elation{P}rompt: Leveraging Prompts to Generate
  Synthetic Data for Zero-Shot Relation Triplet Extraction}. In
  \bibinfo{booktitle}{\emph{Findings of the Association for Computational
  Linguistics: ACL 2022}}. \bibinfo{publisher}{Association for Computational
  Linguistics}, \bibinfo{address}{Dublin, Ireland}, \bibinfo{pages}{45--57}.
\newblock
\urldef\tempurl%
\url{https://doi.org/10.18653/v1/2022.findings-acl.5}
\showDOI{\tempurl}


\bibitem[Cui et~al\mbox{.}(2023)]%
        {cui2023pv2tea}
\bibfield{author}{\bibinfo{person}{Hejie Cui}, \bibinfo{person}{Rongmei Lin},
  \bibinfo{person}{Nasser Zalmout}, \bibinfo{person}{Chenwei Zhang},
  \bibinfo{person}{Jingbo Shang}, \bibinfo{person}{Carl Yang}, {and}
  \bibinfo{person}{Xian Li}.} \bibinfo{year}{2023}\natexlab{}.
\newblock \bibinfo{title}{PV2TEA: Patching Visual Modality to
  Textual-Established Information Extraction}.
\newblock
\newblock
\showeprint[arxiv]{2306.01016}~[cs.CL]


\bibitem[Deng et~al\mbox{.}(2022)]%
        {deng2022ae}
\bibfield{author}{\bibinfo{person}{Zhongfen Deng}, \bibinfo{person}{Wei-Te
  Chen}, \bibinfo{person}{Lei Chen}, {and} \bibinfo{person}{S~Yu Philip}.}
  \bibinfo{year}{2022}\natexlab{}.
\newblock \showarticletitle{AE-smnsMLC: Multi-Label Classification with
  Semantic Matching and Negative Label Sampling for Product Attribute Value
  Extraction}. In \bibinfo{booktitle}{\emph{2022 IEEE International Conference
  on Big Data (Big Data)}}. IEEE, \bibinfo{pages}{1816--1821}.
\newblock


\bibitem[Devlin et~al\mbox{.}(2019)]%
        {devlin-etal-2019-bert}
\bibfield{author}{\bibinfo{person}{Jacob Devlin}, \bibinfo{person}{Ming-Wei
  Chang}, \bibinfo{person}{Kenton Lee}, {and} \bibinfo{person}{Kristina
  Toutanova}.} \bibinfo{year}{2019}\natexlab{}.
\newblock \showarticletitle{{BERT}: Pre-training of Deep Bidirectional
  Transformers for Language Understanding}. In
  \bibinfo{booktitle}{\emph{Proceedings of the 2019 Conference of the North
  {A}merican Chapter of the Association for Computational Linguistics: Human
  Language Technologies, Volume 1 (Long and Short Papers)}}.
  \bibinfo{publisher}{Association for Computational Linguistics},
  \bibinfo{address}{Minneapolis, Minnesota}, \bibinfo{pages}{4171--4186}.
\newblock
\urldef\tempurl%
\url{https://doi.org/10.18653/v1/N19-1423}
\showDOI{\tempurl}


\bibitem[Ding et~al\mbox{.}(2020)]%
        {ding-etal-2020-less}
\bibfield{author}{\bibinfo{person}{Kaize Ding}, \bibinfo{person}{Jianling
  Wang}, \bibinfo{person}{Jundong Li}, \bibinfo{person}{Dingcheng Li}, {and}
  \bibinfo{person}{Huan Liu}.} \bibinfo{year}{2020}\natexlab{}.
\newblock \showarticletitle{Be More with Less: Hypergraph Attention Networks
  for Inductive Text Classification}. In \bibinfo{booktitle}{\emph{Proceedings
  of the 2020 Conference on Empirical Methods in Natural Language Processing
  (EMNLP)}}. \bibinfo{publisher}{Association for Computational Linguistics},
  \bibinfo{address}{Online}, \bibinfo{pages}{4927--4936}.
\newblock
\urldef\tempurl%
\url{https://doi.org/10.18653/v1/2020.emnlp-main.399}
\showDOI{\tempurl}


\bibitem[Fan et~al\mbox{.}(2021)]%
        {fan2021heterogeneous}
\bibfield{author}{\bibinfo{person}{Haoyi Fan}, \bibinfo{person}{Fengbin Zhang},
  \bibinfo{person}{Yuxuan Wei}, \bibinfo{person}{Zuoyong Li},
  \bibinfo{person}{Changqing Zou}, \bibinfo{person}{Yue Gao}, {and}
  \bibinfo{person}{Qionghai Dai}.} \bibinfo{year}{2021}\natexlab{}.
\newblock \showarticletitle{Heterogeneous hypergraph variational autoencoder
  for link prediction}.
\newblock \bibinfo{journal}{\emph{IEEE Transactions on Pattern Analysis and
  Machine Intelligence}} \bibinfo{volume}{44}, \bibinfo{number}{8}
  (\bibinfo{year}{2021}), \bibinfo{pages}{4125--4138}.
\newblock


\bibitem[Fan et~al\mbox{.}(2023)]%
        {fan2023zero}
\bibfield{author}{\bibinfo{person}{Ziwei Fan}, \bibinfo{person}{Zhiwei Liu},
  \bibinfo{person}{Shelby Heinecke}, \bibinfo{person}{Jianguo Zhang},
  \bibinfo{person}{Huan Wang}, \bibinfo{person}{Caiming Xiong}, {and}
  \bibinfo{person}{Philip~S Yu}.} \bibinfo{year}{2023}\natexlab{}.
\newblock \showarticletitle{Zero-shot Item-based Recommendation via Multi-task
  Product Knowledge Graph Pre-Training}.
\newblock \bibinfo{journal}{\emph{arXiv preprint arXiv:2305.07633}}
  (\bibinfo{year}{2023}).
\newblock


\bibitem[Gao et~al\mbox{.}(2023)]%
        {9795251}
\bibfield{author}{\bibinfo{person}{Yue Gao}, \bibinfo{person}{Yifan Feng},
  \bibinfo{person}{Shuyi Ji}, {and} \bibinfo{person}{Rongrong Ji}.}
  \bibinfo{year}{2023}\natexlab{}.
\newblock \showarticletitle{HGNN+: General Hypergraph Neural Networks}.
\newblock \bibinfo{journal}{\emph{IEEE Transactions on Pattern Analysis and
  Machine Intelligence}} \bibinfo{volume}{45}, \bibinfo{number}{3}
  (\bibinfo{year}{2023}), \bibinfo{pages}{3181--3199}.
\newblock
\urldef\tempurl%
\url{https://doi.org/10.1109/TPAMI.2022.3182052}
\showDOI{\tempurl}


\bibitem[Geng et~al\mbox{.}(2022)]%
        {geng2022disentangled}
\bibfield{author}{\bibinfo{person}{Yuxia Geng}, \bibinfo{person}{Jiaoyan Chen},
  \bibinfo{person}{Wen Zhang}, \bibinfo{person}{Yajing Xu},
  \bibinfo{person}{Zhuo Chen}, \bibinfo{person}{Jeff Z.~Pan},
  \bibinfo{person}{Yufeng Huang}, \bibinfo{person}{Feiyu Xiong}, {and}
  \bibinfo{person}{Huajun Chen}.} \bibinfo{year}{2022}\natexlab{}.
\newblock \showarticletitle{Disentangled ontology embedding for zero-shot
  learning}. In \bibinfo{booktitle}{\emph{Proceedings of the 28th ACM SIGKDD
  Conference on Knowledge Discovery and Data Mining}}.
  \bibinfo{pages}{443--453}.
\newblock


\bibitem[Geng et~al\mbox{.}(2023)]%
        {geng2023benchmarking}
\bibfield{author}{\bibinfo{person}{Yuxia Geng}, \bibinfo{person}{Jiaoyan Chen},
  \bibinfo{person}{Xiang Zhuang}, \bibinfo{person}{Zhuo Chen},
  \bibinfo{person}{Jeff~Z Pan}, \bibinfo{person}{Juan Li},
  \bibinfo{person}{Zonggang Yuan}, {and} \bibinfo{person}{Huajun Chen}.}
  \bibinfo{year}{2023}\natexlab{}.
\newblock \showarticletitle{Benchmarking knowledge-driven zero-shot learning}.
\newblock \bibinfo{journal}{\emph{Journal of Web Semantics}}
  \bibinfo{volume}{75} (\bibinfo{year}{2023}), \bibinfo{pages}{100757}.
\newblock


\bibitem[Ghani et~al\mbox{.}(2006)]%
        {10.1145/1147234.1147241}
\bibfield{author}{\bibinfo{person}{Rayid Ghani}, \bibinfo{person}{Katharina
  Probst}, \bibinfo{person}{Yan Liu}, \bibinfo{person}{Marko Krema}, {and}
  \bibinfo{person}{Andrew Fano}.} \bibinfo{year}{2006}\natexlab{}.
\newblock \showarticletitle{Text Mining for Product Attribute Extraction}.
\newblock \bibinfo{journal}{\emph{SIGKDD Explor. Newsl.}} \bibinfo{volume}{8},
  \bibinfo{number}{1} (\bibinfo{date}{jun} \bibinfo{year}{2006}),
  \bibinfo{pages}{41–48}.
\newblock
\showISSN{1931-0145}
\urldef\tempurl%
\url{https://doi.org/10.1145/1147234.1147241}
\showDOI{\tempurl}


\bibitem[Ghosh et~al\mbox{.}(2023)]%
        {ghosh2023d}
\bibfield{author}{\bibinfo{person}{Pushpendu Ghosh}, \bibinfo{person}{Nancy
  Wang}, {and} \bibinfo{person}{Promod Yenigalla}.}
  \bibinfo{year}{2023}\natexlab{}.
\newblock \showarticletitle{D-Extract: Extracting Dimensional Attributes From
  Product Images}. In \bibinfo{booktitle}{\emph{Proceedings of the IEEE/CVF
  Winter Conference on Applications of Computer Vision}}.
  \bibinfo{pages}{3641--3649}.
\newblock


\bibitem[Gong et~al\mbox{.}(2023)]%
        {gong2023knowledge}
\bibfield{author}{\bibinfo{person}{Jiaying Gong}, \bibinfo{person}{Wei-Te
  Chen}, {and} \bibinfo{person}{Hoda Eldardiry}.}
  \bibinfo{year}{2023}\natexlab{}.
\newblock \showarticletitle{Knowledge-Enhanced Multi-Label Few-Shot Product
  Attribute-Value Extraction}.
\newblock \bibinfo{journal}{\emph{arXiv preprint arXiv:2308.08413}}
  (\bibinfo{year}{2023}).
\newblock


\bibitem[Gong and Eldardiry(2021)]%
        {10.1145/3459637.3482403}
\bibfield{author}{\bibinfo{person}{Jiaying Gong} {and} \bibinfo{person}{Hoda
  Eldardiry}.} \bibinfo{year}{2021}\natexlab{}.
\newblock \showarticletitle{Zero-Shot Relation Classification from Side
  Information}. In \bibinfo{booktitle}{\emph{Proceedings of the 30th ACM
  International Conference on Information \& Knowledge Management}} (Virtual
  Event, Queensland, Australia) \emph{(\bibinfo{series}{CIKM '21})}.
  \bibinfo{publisher}{Association for Computing Machinery},
  \bibinfo{address}{New York, NY, USA}, \bibinfo{pages}{576–585}.
\newblock
\showISBNx{9781450384469}
\urldef\tempurl%
\url{https://doi.org/10.1145/3459637.3482403}
\showDOI{\tempurl}


\bibitem[Gong and Eldardiry(2023)]%
        {gong2023promptbased}
\bibfield{author}{\bibinfo{person}{Jiaying Gong} {and} \bibinfo{person}{Hoda
  Eldardiry}.} \bibinfo{year}{2023}\natexlab{}.
\newblock \bibinfo{title}{Prompt-based Zero-shot Relation Extraction with
  Semantic Knowledge Augmentation}.
\newblock
\newblock
\showeprint[arxiv]{2112.04539}~[cs.CL]


\bibitem[Gromann et~al\mbox{.}(2021)]%
        {10.3233/SW-210435}
\bibfield{author}{\bibinfo{person}{Dagmar Gromann}, \bibinfo{person}{Yuxia
  Geng}, \bibinfo{person}{Jiaoyan Chen}, \bibinfo{person}{Zhiquan Ye},
  \bibinfo{person}{Zonggang Yuan}, \bibinfo{person}{Wei Zhang}, {and}
  \bibinfo{person}{Huajun Chen}.} \bibinfo{year}{2021}\natexlab{}.
\newblock \showarticletitle{Explainable Zero-Shot Learning via Attentive Graph
  Convolutional Network and Knowledge Graphs}.
\newblock \bibinfo{journal}{\emph{Semant. Web}} \bibinfo{volume}{12},
  \bibinfo{number}{5} (\bibinfo{date}{jan} \bibinfo{year}{2021}),
  \bibinfo{pages}{741–765}.
\newblock
\showISSN{1570-0844}
\urldef\tempurl%
\url{https://doi.org/10.3233/SW-210435}
\showDOI{\tempurl}


\bibitem[Guzzi and Zitnik(2022)]%
        {guzzi2022editorial}
\bibfield{author}{\bibinfo{person}{Pietro~Hiram Guzzi} {and}
  \bibinfo{person}{Marinka Zitnik}.} \bibinfo{year}{2022}\natexlab{}.
\newblock \showarticletitle{Editorial deep learning and graph embeddings for
  network biology}.
\newblock \bibinfo{journal}{\emph{IEEE/ACM Transactions on Computational
  Biology and Bioinformatics}} \bibinfo{volume}{19}, \bibinfo{number}{2}
  (\bibinfo{year}{2022}), \bibinfo{pages}{653--654}.
\newblock


\bibitem[Han et~al\mbox{.}(2023)]%
        {han2023intra}
\bibfield{author}{\bibinfo{person}{Zhongxuan Han}, \bibinfo{person}{Xiaolin
  Zheng}, \bibinfo{person}{Chaochao Chen}, \bibinfo{person}{Wenjie Cheng},
  {and} \bibinfo{person}{Yang Yao}.} \bibinfo{year}{2023}\natexlab{}.
\newblock \showarticletitle{Intra and Inter Domain HyperGraph Convolutional
  Network for Cross-Domain Recommendation}. In
  \bibinfo{booktitle}{\emph{Proceedings of the ACM Web Conference 2023}}.
  \bibinfo{pages}{449--459}.
\newblock


\bibitem[Hu et~al\mbox{.}(2020)]%
        {hu2020heterogeneous}
\bibfield{author}{\bibinfo{person}{Ziniu Hu}, \bibinfo{person}{Yuxiao Dong},
  \bibinfo{person}{Kuansan Wang}, {and} \bibinfo{person}{Yizhou Sun}.}
  \bibinfo{year}{2020}\natexlab{}.
\newblock \showarticletitle{Heterogeneous graph transformer}. In
  \bibinfo{booktitle}{\emph{Proceedings of the web conference 2020}}.
  \bibinfo{pages}{2704--2710}.
\newblock


\bibitem[Huang et~al\mbox{.}(2022)]%
        {huang2022contexting}
\bibfield{author}{\bibinfo{person}{Yen-Hao Huang}, \bibinfo{person}{Yi-Hsin
  Chen}, {and} \bibinfo{person}{Yi-Shin Chen}.}
  \bibinfo{year}{2022}\natexlab{}.
\newblock \showarticletitle{ConTextING: Granting Document-Wise Contextual
  Embeddings to Graph Neural Networks for Inductive Text Classification}. In
  \bibinfo{booktitle}{\emph{Proceedings of the 29th International Conference on
  Computational Linguistics}}. \bibinfo{pages}{1163--1168}.
\newblock


\bibitem[Jain et~al\mbox{.}(2021)]%
        {jain-etal-2021-learning}
\bibfield{author}{\bibinfo{person}{Mayank Jain}, \bibinfo{person}{Sourangshu
  Bhattacharya}, \bibinfo{person}{Harshit Jain}, \bibinfo{person}{Karimulla
  Shaik}, {and} \bibinfo{person}{Muthusamy Chelliah}.}
  \bibinfo{year}{2021}\natexlab{}.
\newblock \showarticletitle{Learning Cross-Task Attribute - Attribute
  Similarity for Multi-task Attribute-Value Extraction}. In
  \bibinfo{booktitle}{\emph{Proceedings of the 4th Workshop on e-Commerce and
  NLP}}. \bibinfo{publisher}{Association for Computational Linguistics},
  \bibinfo{address}{Online}, \bibinfo{pages}{79--87}.
\newblock
\urldef\tempurl%
\url{https://doi.org/10.18653/v1/2021.ecnlp-1.10}
\showDOI{\tempurl}


\bibitem[Khan et~al\mbox{.}(2023)]%
        {khan2023heterogeneous}
\bibfield{author}{\bibinfo{person}{Bilal Khan}, \bibinfo{person}{Jia Wu},
  \bibinfo{person}{Jian Yang}, {and} \bibinfo{person}{Xiaoxiao Ma}.}
  \bibinfo{year}{2023}\natexlab{}.
\newblock \showarticletitle{Heterogeneous Hypergraph Neural Network for Social
  Recommendation using Attention Network}.
\newblock \bibinfo{journal}{\emph{ACM Transactions on Recommender Systems}}
  (\bibinfo{year}{2023}).
\newblock


\bibitem[Kingma and Welling(2013)]%
        {kingma2013auto}
\bibfield{author}{\bibinfo{person}{Diederik~P Kingma} {and}
  \bibinfo{person}{Max Welling}.} \bibinfo{year}{2013}\natexlab{}.
\newblock \showarticletitle{Auto-encoding variational bayes}.
\newblock \bibinfo{journal}{\emph{arXiv preprint arXiv:1312.6114}}
  (\bibinfo{year}{2013}).
\newblock


\bibitem[Lewis et~al\mbox{.}(2019)]%
        {lewis2019bart}
\bibfield{author}{\bibinfo{person}{Mike Lewis}, \bibinfo{person}{Yinhan Liu},
  \bibinfo{person}{Naman Goyal}, \bibinfo{person}{Marjan Ghazvininejad},
  \bibinfo{person}{Abdelrahman Mohamed}, \bibinfo{person}{Omer Levy},
  \bibinfo{person}{Ves Stoyanov}, {and} \bibinfo{person}{Luke Zettlemoyer}.}
  \bibinfo{year}{2019}\natexlab{}.
\newblock \showarticletitle{Bart: Denoising sequence-to-sequence pre-training
  for natural language generation, translation, and comprehension}.
\newblock \bibinfo{journal}{\emph{arXiv preprint arXiv:1910.13461}}
  (\bibinfo{year}{2019}).
\newblock


\bibitem[Li et~al\mbox{.}(2023)]%
        {li2023hypergraph}
\bibfield{author}{\bibinfo{person}{Mengran Li}, \bibinfo{person}{Yong Zhang},
  \bibinfo{person}{Xiaoyong Li}, \bibinfo{person}{Yuchen Zhang}, {and}
  \bibinfo{person}{Baocai Yin}.} \bibinfo{year}{2023}\natexlab{}.
\newblock \showarticletitle{Hypergraph transformer neural networks}.
\newblock \bibinfo{journal}{\emph{ACM Transactions on Knowledge Discovery from
  Data}} \bibinfo{volume}{17}, \bibinfo{number}{5} (\bibinfo{year}{2023}),
  \bibinfo{pages}{1--22}.
\newblock


\bibitem[Li et~al\mbox{.}(2022a)]%
        {li2022hmgcl}
\bibfield{author}{\bibinfo{person}{Yongkang Li}, \bibinfo{person}{Zipei Fan},
  \bibinfo{person}{Du Yin}, \bibinfo{person}{Renhe Jiang},
  \bibinfo{person}{Jinliang Deng}, {and} \bibinfo{person}{Xuan Song}.}
  \bibinfo{year}{2022}\natexlab{a}.
\newblock \showarticletitle{HMGCL: Heterogeneous multigraph contrastive
  learning for LBSN friend recommendation}.
\newblock \bibinfo{journal}{\emph{World Wide Web}} (\bibinfo{year}{2022}),
  \bibinfo{pages}{1--24}.
\newblock


\bibitem[Li et~al\mbox{.}(2022b)]%
        {li2022heterogeneous}
\bibfield{author}{\bibinfo{person}{Yongkang Li}, \bibinfo{person}{Zipei Fan},
  \bibinfo{person}{Jixiao Zhang}, \bibinfo{person}{Dengheng Shi},
  \bibinfo{person}{Tianqi Xu}, \bibinfo{person}{Du Yin},
  \bibinfo{person}{Jinliang Deng}, {and} \bibinfo{person}{Xuan Song}.}
  \bibinfo{year}{2022}\natexlab{b}.
\newblock \showarticletitle{Heterogeneous Hypergraph Neural Network for Friend
  Recommendation with Human Mobility}. In \bibinfo{booktitle}{\emph{Proceedings
  of the 31st ACM International Conference on Information \& Knowledge
  Management}}. \bibinfo{pages}{4209--4213}.
\newblock


\bibitem[Lin et~al\mbox{.}(2021)]%
        {lin2021pam}
\bibfield{author}{\bibinfo{person}{Rongmei Lin}, \bibinfo{person}{Xiang He},
  \bibinfo{person}{Jie Feng}, \bibinfo{person}{Nasser Zalmout},
  \bibinfo{person}{Yan Liang}, \bibinfo{person}{Li Xiong}, {and}
  \bibinfo{person}{Xin~Luna Dong}.} \bibinfo{year}{2021}\natexlab{}.
\newblock \showarticletitle{PAM: understanding product images in cross product
  category attribute extraction}. In \bibinfo{booktitle}{\emph{Proceedings of
  the 27th ACM SIGKDD Conference on Knowledge Discovery \& Data Mining}}.
  \bibinfo{pages}{3262--3270}.
\newblock


\bibitem[Liu et~al\mbox{.}(2022b)]%
        {liu-etal-2022-pre}
\bibfield{author}{\bibinfo{person}{Fangchao Liu}, \bibinfo{person}{Hongyu Lin},
  \bibinfo{person}{Xianpei Han}, \bibinfo{person}{Boxi Cao}, {and}
  \bibinfo{person}{Le Sun}.} \bibinfo{year}{2022}\natexlab{b}.
\newblock \showarticletitle{Pre-training to Match for Unified Low-shot Relation
  Extraction}. In \bibinfo{booktitle}{\emph{Proceedings of the 60th Annual
  Meeting of the Association for Computational Linguistics (Volume 1: Long
  Papers)}}. \bibinfo{publisher}{Association for Computational Linguistics},
  \bibinfo{address}{Dublin, Ireland}, \bibinfo{pages}{5785--5795}.
\newblock
\urldef\tempurl%
\url{https://doi.org/10.18653/v1/2022.acl-long.397}
\showDOI{\tempurl}


\bibitem[Liu et~al\mbox{.}(2023a)]%
        {liu2023JDsearch}
\bibfield{author}{\bibinfo{person}{Jiongnan Liu}, \bibinfo{person}{Zhicheng
  Dou}, \bibinfo{person}{Guoyu Tang}, {and} \bibinfo{person}{Sulong Xu}.}
  \bibinfo{year}{2023}\natexlab{a}.
\newblock \showarticletitle{JDsearch: A Personalized Product Search Dataset
  with Real Queries and Full Interactions}. In
  \bibinfo{booktitle}{\emph{Proceedings of the {SIGIR} 2023}}.
  \bibinfo{publisher}{{ACM}}.
\newblock
\urldef\tempurl%
\url{https://doi.org/10.1145/3539618.3591900}
\showDOI{\tempurl}


\bibitem[Liu et~al\mbox{.}(2022a)]%
        {liu2022hypergraph}
\bibfield{author}{\bibinfo{person}{Jingquan Liu}, \bibinfo{person}{Xiaoyong
  Du}, \bibinfo{person}{Yuanzhe Li}, {and} \bibinfo{person}{Weidong Hu}.}
  \bibinfo{year}{2022}\natexlab{a}.
\newblock \showarticletitle{Hypergraph Variational Autoencoder for Multimodal
  Semi-supervised Representation Learning}. In
  \bibinfo{booktitle}{\emph{Artificial Neural Networks and Machine
  Learning--ICANN 2022: 31st International Conference on Artificial Neural
  Networks, Bristol, UK, September 6--9, 2022, Proceedings; Part IV}}.
  Springer, \bibinfo{pages}{395--406}.
\newblock


\bibitem[Liu et~al\mbox{.}(2023c)]%
        {liu2023meta}
\bibfield{author}{\bibinfo{person}{Jie Liu}, \bibinfo{person}{Lingyun Song},
  \bibinfo{person}{Guangtao Wang}, {and} \bibinfo{person}{Xuequn Shang}.}
  \bibinfo{year}{2023}\natexlab{c}.
\newblock \showarticletitle{Meta-HGT: Metapath-aware HyperGraph Transformer for
  heterogeneous information network embedding}.
\newblock \bibinfo{journal}{\emph{Neural Networks}}  \bibinfo{volume}{157}
  (\bibinfo{year}{2023}), \bibinfo{pages}{65--76}.
\newblock


\bibitem[Liu et~al\mbox{.}(2023d)]%
        {liu2023boosting}
\bibfield{author}{\bibinfo{person}{Mengyin Liu}, \bibinfo{person}{Chao Zhu},
  \bibinfo{person}{Hongyu Gao}, \bibinfo{person}{Weibo Gu},
  \bibinfo{person}{Hongfa Wang}, \bibinfo{person}{Wei Liu}, {and}
  \bibinfo{person}{Xu cheng Yin}.} \bibinfo{year}{2023}\natexlab{d}.
\newblock \bibinfo{title}{Boosting Multi-Modal E-commerce Attribute Value
  Extraction via Unified Learning Scheme and Dynamic Range Minimization}.
\newblock
\newblock
\showeprint[arxiv]{2207.07278}~[cs.CV]


\bibitem[Liu et~al\mbox{.}(2023b)]%
        {liu2023multimodal}
\bibfield{author}{\bibinfo{person}{Shilei Liu}, \bibinfo{person}{Lin Li},
  \bibinfo{person}{Jun Song}, \bibinfo{person}{Yonghua Yang}, {and}
  \bibinfo{person}{Xiaoyi Zeng}.} \bibinfo{year}{2023}\natexlab{b}.
\newblock \showarticletitle{Multimodal Pre-Training with Self-Distillation for
  Product Understanding in E-Commerce}. In
  \bibinfo{booktitle}{\emph{Proceedings of the Sixteenth ACM International
  Conference on Web Search and Data Mining}}. \bibinfo{pages}{1039--1047}.
\newblock


\bibitem[Ma et~al\mbox{.}(2020)]%
        {ma2020transductive}
\bibfield{author}{\bibinfo{person}{Yuqing Ma}, \bibinfo{person}{Shihao Bai},
  \bibinfo{person}{Shan An}, \bibinfo{person}{Wei Liu}, \bibinfo{person}{Aishan
  Liu}, \bibinfo{person}{Xiantong Zhen}, {and} \bibinfo{person}{Xianglong
  Liu}.} \bibinfo{year}{2020}\natexlab{}.
\newblock \showarticletitle{Transductive Relation-Propagation Network for
  Few-shot Learning.}. In \bibinfo{booktitle}{\emph{IJCAI}},
  Vol.~\bibinfo{volume}{20}. \bibinfo{pages}{804--810}.
\newblock


\bibitem[Milano et~al\mbox{.}(2022)]%
        {milano2022challenges}
\bibfield{author}{\bibinfo{person}{Marianna Milano}, \bibinfo{person}{Giuseppe
  Agapito}, {and} \bibinfo{person}{Mario Cannataro}.}
  \bibinfo{year}{2022}\natexlab{}.
\newblock \showarticletitle{Challenges and limitations of biological network
  analysis}.
\newblock \bibinfo{journal}{\emph{BioTech}} \bibinfo{volume}{11},
  \bibinfo{number}{3} (\bibinfo{year}{2022}), \bibinfo{pages}{24}.
\newblock


\bibitem[Mirza and Osindero(2014)]%
        {mirza2014conditional}
\bibfield{author}{\bibinfo{person}{Mehdi Mirza} {and} \bibinfo{person}{Simon
  Osindero}.} \bibinfo{year}{2014}\natexlab{}.
\newblock \showarticletitle{Conditional generative adversarial nets}.
\newblock \bibinfo{journal}{\emph{arXiv preprint arXiv:1411.1784}}
  (\bibinfo{year}{2014}).
\newblock


\bibitem[Ni et~al\mbox{.}(2019)]%
        {ni2019justifying}
\bibfield{author}{\bibinfo{person}{Jianmo Ni}, \bibinfo{person}{Jiacheng Li},
  {and} \bibinfo{person}{Julian McAuley}.} \bibinfo{year}{2019}\natexlab{}.
\newblock \showarticletitle{Justifying recommendations using distantly-labeled
  reviews and fine-grained aspects}. In \bibinfo{booktitle}{\emph{Proceedings
  of the 2019 conference on empirical methods in natural language processing
  and the 9th international joint conference on natural language processing
  (EMNLP-IJCNLP)}}. \bibinfo{pages}{188--197}.
\newblock


\bibitem[Pourpanah et~al\mbox{.}(2023)]%
        {9832795}
\bibfield{author}{\bibinfo{person}{Farhad Pourpanah}, \bibinfo{person}{Moloud
  Abdar}, \bibinfo{person}{Yuxuan Luo}, \bibinfo{person}{Xinlei Zhou},
  \bibinfo{person}{Ran Wang}, \bibinfo{person}{Chee~Peng Lim},
  \bibinfo{person}{Xi-Zhao Wang}, {and} \bibinfo{person}{Q.~M.~Jonathan Wu}.}
  \bibinfo{year}{2023}\natexlab{}.
\newblock \showarticletitle{A Review of Generalized Zero-Shot Learning
  Methods}.
\newblock \bibinfo{journal}{\emph{IEEE Transactions on Pattern Analysis and
  Machine Intelligence}} \bibinfo{volume}{45}, \bibinfo{number}{4}
  (\bibinfo{year}{2023}), \bibinfo{pages}{4051--4070}.
\newblock
\urldef\tempurl%
\url{https://doi.org/10.1109/TPAMI.2022.3191696}
\showDOI{\tempurl}


\bibitem[Putthividhya and Hu(2011)]%
        {10.5555/2145432.2145598}
\bibfield{author}{\bibinfo{person}{Duangmanee~(Pew) Putthividhya} {and}
  \bibinfo{person}{Junling Hu}.} \bibinfo{year}{2011}\natexlab{}.
\newblock \showarticletitle{Bootstrapped Named Entity Recognition for Product
  Attribute Extraction}. In \bibinfo{booktitle}{\emph{Proceedings of the
  Conference on Empirical Methods in Natural Language Processing}} (Edinburgh,
  United Kingdom) \emph{(\bibinfo{series}{EMNLP '11})}.
  \bibinfo{publisher}{Association for Computational Linguistics},
  \bibinfo{address}{USA}, \bibinfo{pages}{1557–1567}.
\newblock
\showISBNx{9781937284114}


\bibitem[Radford et~al\mbox{.}(2019)]%
        {radford2019language}
\bibfield{author}{\bibinfo{person}{Alec Radford}, \bibinfo{person}{Jeffrey Wu},
  \bibinfo{person}{Rewon Child}, \bibinfo{person}{David Luan},
  \bibinfo{person}{Dario Amodei}, \bibinfo{person}{Ilya Sutskever},
  {et~al\mbox{.}}} \bibinfo{year}{2019}\natexlab{}.
\newblock \showarticletitle{Language models are unsupervised multitask
  learners}.
\newblock \bibinfo{journal}{\emph{OpenAI blog}} \bibinfo{volume}{1},
  \bibinfo{number}{8} (\bibinfo{year}{2019}), \bibinfo{pages}{9}.
\newblock


\bibitem[Raffel et~al\mbox{.}(2020)]%
        {raffel2020exploring}
\bibfield{author}{\bibinfo{person}{Colin Raffel}, \bibinfo{person}{Noam
  Shazeer}, \bibinfo{person}{Adam Roberts}, \bibinfo{person}{Katherine Lee},
  \bibinfo{person}{Sharan Narang}, \bibinfo{person}{Michael Matena},
  \bibinfo{person}{Yanqi Zhou}, \bibinfo{person}{Wei Li}, {and}
  \bibinfo{person}{Peter~J Liu}.} \bibinfo{year}{2020}\natexlab{}.
\newblock \showarticletitle{Exploring the limits of transfer learning with a
  unified text-to-text transformer}.
\newblock \bibinfo{journal}{\emph{The Journal of Machine Learning Research}}
  \bibinfo{volume}{21}, \bibinfo{number}{1} (\bibinfo{year}{2020}),
  \bibinfo{pages}{5485--5551}.
\newblock


\bibitem[Ragesh et~al\mbox{.}(2021)]%
        {ragesh2021hetegcn}
\bibfield{author}{\bibinfo{person}{Rahul Ragesh}, \bibinfo{person}{Sundararajan
  Sellamanickam}, \bibinfo{person}{Arun Iyer}, \bibinfo{person}{Ramakrishna
  Bairi}, {and} \bibinfo{person}{Vijay Lingam}.}
  \bibinfo{year}{2021}\natexlab{}.
\newblock \showarticletitle{Hetegcn: heterogeneous graph convolutional networks
  for text classification}. In \bibinfo{booktitle}{\emph{Proceedings of the
  14th ACM international conference on web search and data mining}}.
  \bibinfo{pages}{860--868}.
\newblock


\bibitem[Rezk et~al\mbox{.}(2019)]%
        {8731553}
\bibfield{author}{\bibinfo{person}{Martin Rezk}, \bibinfo{person}{Laura
  Alonso~Alemany}, \bibinfo{person}{Lasguido Nio}, {and} \bibinfo{person}{Ted
  Zhang}.} \bibinfo{year}{2019}\natexlab{}.
\newblock \showarticletitle{Accurate Product Attribute Extraction on the
  Field}. In \bibinfo{booktitle}{\emph{2019 IEEE 35th International Conference
  on Data Engineering (ICDE)}}. \bibinfo{pages}{1862--1873}.
\newblock
\urldef\tempurl%
\url{https://doi.org/10.1109/ICDE.2019.00202}
\showDOI{\tempurl}


\bibitem[Roy et~al\mbox{.}(2021)]%
        {roy2021attribute}
\bibfield{author}{\bibinfo{person}{Kalyani Roy}, \bibinfo{person}{Pawan Goyal},
  {and} \bibinfo{person}{Manish Pandey}.} \bibinfo{year}{2021}\natexlab{}.
\newblock \showarticletitle{Attribute value generation from product title using
  language models}. In \bibinfo{booktitle}{\emph{Proceedings of The 4th
  Workshop on e-Commerce and NLP}}. \bibinfo{pages}{13--17}.
\newblock


\bibitem[Roy et~al\mbox{.}(2022)]%
        {roy2022exploring}
\bibfield{author}{\bibinfo{person}{Kalyani Roy}, \bibinfo{person}{Tapas Nayak},
  {and} \bibinfo{person}{Pawan Goyal}.} \bibinfo{year}{2022}\natexlab{}.
\newblock \showarticletitle{Exploring Generative Models for Joint Attribute
  Value Extraction from Product Titles}.
\newblock \bibinfo{journal}{\emph{arXiv preprint arXiv:2208.07130}}
  (\bibinfo{year}{2022}).
\newblock


\bibitem[Sainz et~al\mbox{.}(2021)]%
        {sainz-etal-2021-label}
\bibfield{author}{\bibinfo{person}{Oscar Sainz}, \bibinfo{person}{Oier Lopez~de
  Lacalle}, \bibinfo{person}{Gorka Labaka}, \bibinfo{person}{Ander Barrena},
  {and} \bibinfo{person}{Eneko Agirre}.} \bibinfo{year}{2021}\natexlab{}.
\newblock \showarticletitle{Label Verbalization and Entailment for Effective
  Zero and Few-Shot Relation Extraction}. In
  \bibinfo{booktitle}{\emph{Proceedings of the 2021 Conference on Empirical
  Methods in Natural Language Processing}}. \bibinfo{publisher}{Association for
  Computational Linguistics}, \bibinfo{address}{Online and Punta Cana,
  Dominican Republic}, \bibinfo{pages}{1199--1212}.
\newblock
\urldef\tempurl%
\url{https://doi.org/10.18653/v1/2021.emnlp-main.92}
\showDOI{\tempurl}


\bibitem[Scarselli et~al\mbox{.}(2008)]%
        {scarselli2008graph}
\bibfield{author}{\bibinfo{person}{Franco Scarselli}, \bibinfo{person}{Marco
  Gori}, \bibinfo{person}{Ah~Chung Tsoi}, \bibinfo{person}{Markus
  Hagenbuchner}, {and} \bibinfo{person}{Gabriele Monfardini}.}
  \bibinfo{year}{2008}\natexlab{}.
\newblock \showarticletitle{The graph neural network model}.
\newblock \bibinfo{journal}{\emph{IEEE transactions on neural networks}}
  \bibinfo{volume}{20}, \bibinfo{number}{1} (\bibinfo{year}{2008}),
  \bibinfo{pages}{61--80}.
\newblock


\bibitem[Shinzato et~al\mbox{.}(2022)]%
        {shinzato2022simple}
\bibfield{author}{\bibinfo{person}{Keiji Shinzato}, \bibinfo{person}{Naoki
  Yoshinaga}, \bibinfo{person}{Yandi Xia}, {and} \bibinfo{person}{Wei-Te
  Chen}.} \bibinfo{year}{2022}\natexlab{}.
\newblock \showarticletitle{Simple and Effective Knowledge-Driven Query
  Expansion for QA-Based Product Attribute Extraction}. In
  \bibinfo{booktitle}{\emph{Proceedings of the 60th Annual Meeting of the
  Association for Computational Linguistics (Volume 2: Short Papers)}}.
  \bibinfo{pages}{227--234}.
\newblock


\bibitem[Shinzato et~al\mbox{.}(2023)]%
        {shinzato2023unified}
\bibfield{author}{\bibinfo{person}{Keiji Shinzato}, \bibinfo{person}{Naoki
  Yoshinaga}, \bibinfo{person}{Yandi Xia}, {and} \bibinfo{person}{Wei-Te
  Chen}.} \bibinfo{year}{2023}\natexlab{}.
\newblock \showarticletitle{A Unified Generative Approach to Product
  Attribute-Value Identification}.
\newblock \bibinfo{journal}{\emph{arXiv preprint arXiv:2306.05605}}
  (\bibinfo{year}{2023}).
\newblock


\bibitem[Sun et~al\mbox{.}(2021)]%
        {sun2021heterogeneous}
\bibfield{author}{\bibinfo{person}{Xiangguo Sun}, \bibinfo{person}{Hongzhi
  Yin}, \bibinfo{person}{Bo Liu}, \bibinfo{person}{Hongxu Chen},
  \bibinfo{person}{Jiuxin Cao}, \bibinfo{person}{Yingxia Shao}, {and}
  \bibinfo{person}{Nguyen~Quoc Viet~Hung}.} \bibinfo{year}{2021}\natexlab{}.
\newblock \showarticletitle{Heterogeneous hypergraph embedding for graph
  classification}. In \bibinfo{booktitle}{\emph{Proceedings of the 14th ACM
  international conference on web search and data mining}}.
  \bibinfo{pages}{725--733}.
\newblock


\bibitem[Wang et~al\mbox{.}(2023)]%
        {wang2023mpkgac}
\bibfield{author}{\bibinfo{person}{Kai Wang}, \bibinfo{person}{Jianzhi Shao},
  \bibinfo{person}{Tao Zhang}, \bibinfo{person}{Qijin Chen}, {and}
  \bibinfo{person}{Chengfu Huo}.} \bibinfo{year}{2023}\natexlab{}.
\newblock \showarticletitle{MPKGAC: Multimodal Product Attribute Completion in
  E-commerce}. In \bibinfo{booktitle}{\emph{Companion Proceedings of the ACM
  Web Conference 2023}}. \bibinfo{pages}{336--340}.
\newblock


\bibitem[Wang et~al\mbox{.}(2020)]%
        {wang2020learning}
\bibfield{author}{\bibinfo{person}{Qifan Wang}, \bibinfo{person}{Li Yang},
  \bibinfo{person}{Bhargav Kanagal}, \bibinfo{person}{Sumit Sanghai},
  \bibinfo{person}{D Sivakumar}, \bibinfo{person}{Bin Shu},
  \bibinfo{person}{Zac Yu}, {and} \bibinfo{person}{Jon Elsas}.}
  \bibinfo{year}{2020}\natexlab{}.
\newblock \showarticletitle{Learning to extract attribute value from product
  via question answering: A multi-task approach}. In
  \bibinfo{booktitle}{\emph{Proceedings of the 26th ACM SIGKDD International
  Conference on Knowledge Discovery \& Data Mining}}. \bibinfo{pages}{47--55}.
\newblock


\bibitem[Wang et~al\mbox{.}(2022)]%
        {wang2022smartave}
\bibfield{author}{\bibinfo{person}{Qifan Wang}, \bibinfo{person}{Li Yang},
  \bibinfo{person}{Jingang Wang}, \bibinfo{person}{Jitin Krishnan},
  \bibinfo{person}{Bo Dai}, \bibinfo{person}{Sinong Wang},
  \bibinfo{person}{Zenglin Xu}, \bibinfo{person}{Madian Khabsa}, {and}
  \bibinfo{person}{Hao Ma}.} \bibinfo{year}{2022}\natexlab{}.
\newblock \showarticletitle{SMARTAVE: Structured Multimodal Transformer for
  Product Attribute Value Extraction}. In \bibinfo{booktitle}{\emph{Findings of
  the Association for Computational Linguistics: EMNLP 2022}}.
  \bibinfo{pages}{263--276}.
\newblock


\bibitem[Wang et~al\mbox{.}(2019)]%
        {wang2019heterogeneous}
\bibfield{author}{\bibinfo{person}{Xiao Wang}, \bibinfo{person}{Houye Ji},
  \bibinfo{person}{Chuan Shi}, \bibinfo{person}{Bai Wang},
  \bibinfo{person}{Yanfang Ye}, \bibinfo{person}{Peng Cui}, {and}
  \bibinfo{person}{Philip~S Yu}.} \bibinfo{year}{2019}\natexlab{}.
\newblock \showarticletitle{Heterogeneous graph attention network}. In
  \bibinfo{booktitle}{\emph{The world wide web conference}}.
  \bibinfo{pages}{2022--2032}.
\newblock


\bibitem[Wei et~al\mbox{.}(2023)]%
        {wei2023dual}
\bibfield{author}{\bibinfo{person}{Xuemei Wei}, \bibinfo{person}{Yezheng Liu},
  \bibinfo{person}{Jianshan Sun}, \bibinfo{person}{Yuanchun Jiang},
  \bibinfo{person}{Qifeng Tang}, {and} \bibinfo{person}{Kun Yuan}.}
  \bibinfo{year}{2023}\natexlab{}.
\newblock \showarticletitle{Dual subgraph-based graph neural network for
  friendship prediction in location-based social networks}.
\newblock \bibinfo{journal}{\emph{ACM Transactions on Knowledge Discovery from
  Data}} \bibinfo{volume}{17}, \bibinfo{number}{3} (\bibinfo{year}{2023}),
  \bibinfo{pages}{1--28}.
\newblock


\bibitem[Wenping et~al\mbox{.}(2022)]%
        {10010428}
\bibfield{author}{\bibinfo{person}{Zheng Wenping}, \bibinfo{person}{Liu
  Meilin}, {and} \bibinfo{person}{Liang Jiye}.}
  \bibinfo{year}{2022}\natexlab{}.
\newblock \showarticletitle{Hypergraphs: Concepts, Applications and Analysis}.
  In \bibinfo{booktitle}{\emph{2022 IEEE 13th International Symposium on
  Parallel Architectures, Algorithms and Programming (PAAP)}}.
  \bibinfo{pages}{1--6}.
\newblock
\urldef\tempurl%
\url{https://doi.org/10.1109/PAAP56126.2022.10010428}
\showDOI{\tempurl}


\bibitem[Wong et~al\mbox{.}(2009)]%
        {34460}
\bibfield{author}{\bibinfo{person}{Yuk~Wah Wong}, \bibinfo{person}{Dominic
  Widdows}, \bibinfo{person}{Tom Lokovic}, {and} \bibinfo{person}{Kamal
  Nigam}.} \bibinfo{year}{2009}\natexlab{}.
\newblock \showarticletitle{Scalable Attribute-Value Extraction from
  Semi-Structured Text}. In \bibinfo{booktitle}{\emph{ICDM Workshop on
  Large-scale Data Mining: Theory and Applications}}.
\newblock
\urldef\tempurl%
\url{http://www.computer.org/portal/web/csdl/doi/10.1109/ICDMW.2009.81}
\showURL{%
\tempurl}


\bibitem[Wu et~al\mbox{.}(2022)]%
        {wu2022hypergraph}
\bibfield{author}{\bibinfo{person}{Hanrui Wu}, \bibinfo{person}{Yuguang Yan},
  {and} \bibinfo{person}{Michael Kwok-Po Ng}.} \bibinfo{year}{2022}\natexlab{}.
\newblock \showarticletitle{Hypergraph collaborative network on vertices and
  hyperedges}.
\newblock \bibinfo{journal}{\emph{IEEE Transactions on Pattern Analysis and
  Machine Intelligence}} \bibinfo{volume}{45}, \bibinfo{number}{3}
  (\bibinfo{year}{2022}), \bibinfo{pages}{3245--3258}.
\newblock


\bibitem[Xu et~al\mbox{.}(2020)]%
        {xu2020inductive}
\bibfield{author}{\bibinfo{person}{Da Xu}, \bibinfo{person}{Chuanwei Ruan},
  \bibinfo{person}{Evren Korpeoglu}, \bibinfo{person}{Sushant Kumar}, {and}
  \bibinfo{person}{Kannan Achan}.} \bibinfo{year}{2020}\natexlab{}.
\newblock \showarticletitle{Inductive representation learning on temporal
  graphs}.
\newblock \bibinfo{journal}{\emph{arXiv preprint arXiv:2002.07962}}
  (\bibinfo{year}{2020}).
\newblock


\bibitem[Xu et~al\mbox{.}(2019)]%
        {xu2019scaling}
\bibfield{author}{\bibinfo{person}{Huimin Xu}, \bibinfo{person}{Wenting Wang},
  \bibinfo{person}{Xinnian Mao}, \bibinfo{person}{Xinyu Jiang}, {and}
  \bibinfo{person}{Man Lan}.} \bibinfo{year}{2019}\natexlab{}.
\newblock \showarticletitle{Scaling up open tagging from tens to thousands:
  Comprehension empowered attribute value extraction from product title}. In
  \bibinfo{booktitle}{\emph{Proceedings of the 57th Annual Meeting of the
  Association for Computational Linguistics}}. \bibinfo{pages}{5214--5223}.
\newblock


\bibitem[Xu et~al\mbox{.}(2023b)]%
        {xu2023towards}
\bibfield{author}{\bibinfo{person}{Liyan Xu}, \bibinfo{person}{Chenwei Zhang},
  \bibinfo{person}{Xian Li}, \bibinfo{person}{Jingbo Shang}, {and}
  \bibinfo{person}{Jinho~D Choi}.} \bibinfo{year}{2023}\natexlab{b}.
\newblock \showarticletitle{Towards Open-World Product Attribute Mining: A
  Lightly-Supervised Approach}.
\newblock \bibinfo{journal}{\emph{arXiv preprint arXiv:2305.18350}}
  (\bibinfo{year}{2023}).
\newblock


\bibitem[Xu et~al\mbox{.}(2023a)]%
        {xu2023correlative}
\bibfield{author}{\bibinfo{person}{Zixuan Xu}, \bibinfo{person}{Penghui Wei},
  \bibinfo{person}{Shaoguo Liu}, \bibinfo{person}{Weimin Zhang},
  \bibinfo{person}{Liang Wang}, {and} \bibinfo{person}{Bo Zheng}.}
  \bibinfo{year}{2023}\natexlab{a}.
\newblock \showarticletitle{Correlative Preference Transfer with Hierarchical
  Hypergraph Network for Multi-Domain Recommendation}. In
  \bibinfo{booktitle}{\emph{Proceedings of the ACM Web Conference 2023}}.
  \bibinfo{pages}{983--991}.
\newblock


\bibitem[Yadati et~al\mbox{.}(2019)]%
        {yadati2019hypergcn}
\bibfield{author}{\bibinfo{person}{Naganand Yadati}, \bibinfo{person}{Madhav
  Nimishakavi}, \bibinfo{person}{Prateek Yadav}, \bibinfo{person}{Vikram
  Nitin}, \bibinfo{person}{Anand Louis}, {and} \bibinfo{person}{Partha
  Talukdar}.} \bibinfo{year}{2019}\natexlab{}.
\newblock \showarticletitle{Hypergcn: A new method for training graph
  convolutional networks on hypergraphs}.
\newblock \bibinfo{journal}{\emph{Advances in neural information processing
  systems}}  \bibinfo{volume}{32} (\bibinfo{year}{2019}).
\newblock


\bibitem[Yan et~al\mbox{.}(2021)]%
        {yan-etal-2021-adatag}
\bibfield{author}{\bibinfo{person}{Jun Yan}, \bibinfo{person}{Nasser Zalmout},
  \bibinfo{person}{Yan Liang}, \bibinfo{person}{Christan Grant},
  \bibinfo{person}{Xiang Ren}, {and} \bibinfo{person}{Xin~Luna Dong}.}
  \bibinfo{year}{2021}\natexlab{}.
\newblock \showarticletitle{{A}da{T}ag: Multi-Attribute Value Extraction from
  Product Profiles with Adaptive Decoding}. In
  \bibinfo{booktitle}{\emph{Proceedings of the 59th Annual Meeting of the
  Association for Computational Linguistics and the 11th International Joint
  Conference on Natural Language Processing (Volume 1: Long Papers)}}.
  \bibinfo{publisher}{Association for Computational Linguistics},
  \bibinfo{address}{Online}, \bibinfo{pages}{4694--4705}.
\newblock
\urldef\tempurl%
\url{https://doi.org/10.18653/v1/2021.acl-long.362}
\showDOI{\tempurl}


\bibitem[Yang et~al\mbox{.}(2022)]%
        {yang2022mave}
\bibfield{author}{\bibinfo{person}{Li Yang}, \bibinfo{person}{Qifan Wang},
  \bibinfo{person}{Zac Yu}, \bibinfo{person}{Anand Kulkarni},
  \bibinfo{person}{Sumit Sanghai}, \bibinfo{person}{Bin Shu},
  \bibinfo{person}{Jon Elsas}, {and} \bibinfo{person}{Bhargav Kanagal}.}
  \bibinfo{year}{2022}\natexlab{}.
\newblock \showarticletitle{MAVE: A product dataset for multi-source attribute
  value extraction}. In \bibinfo{booktitle}{\emph{Proceedings of the fifteenth
  ACM international conference on web search and data mining}}.
  \bibinfo{pages}{1256--1265}.
\newblock


\bibitem[Zhang et~al\mbox{.}(2022a)]%
        {zhang2022learnable}
\bibfield{author}{\bibinfo{person}{Jiying Zhang}, \bibinfo{person}{Yuzhao
  Chen}, \bibinfo{person}{Xi Xiao}, \bibinfo{person}{Runiu Lu}, {and}
  \bibinfo{person}{Shu-Tao Xia}.} \bibinfo{year}{2022}\natexlab{a}.
\newblock \showarticletitle{Learnable hypergraph laplacian for hypergraph
  learning}. In \bibinfo{booktitle}{\emph{ICASSP 2022-2022 IEEE International
  Conference on Acoustics, Speech and Signal Processing (ICASSP)}}. IEEE,
  \bibinfo{pages}{4503--4507}.
\newblock


\bibitem[Zhang et~al\mbox{.}(2022b)]%
        {zhang2022hypergraph}
\bibfield{author}{\bibinfo{person}{Liyan Zhang}, \bibinfo{person}{Jingfeng
  Guo}, \bibinfo{person}{Jiazheng Wang}, \bibinfo{person}{Jing Wang},
  \bibinfo{person}{Shanshan Li}, {and} \bibinfo{person}{Chunying Zhang}.}
  \bibinfo{year}{2022}\natexlab{b}.
\newblock \showarticletitle{Hypergraph and uncertain hypergraph representation
  learning theory and methods}.
\newblock \bibinfo{journal}{\emph{Mathematics}} \bibinfo{volume}{10},
  \bibinfo{number}{11} (\bibinfo{year}{2022}), \bibinfo{pages}{1921}.
\newblock


\bibitem[Zhang et~al\mbox{.}(2022c)]%
        {10.1145/3485447.3512035}
\bibfield{author}{\bibinfo{person}{Xinyang Zhang}, \bibinfo{person}{Chenwei
  Zhang}, \bibinfo{person}{Xian Li}, \bibinfo{person}{Xin~Luna Dong},
  \bibinfo{person}{Jingbo Shang}, \bibinfo{person}{Christos Faloutsos}, {and}
  \bibinfo{person}{Jiawei Han}.} \bibinfo{year}{2022}\natexlab{c}.
\newblock \showarticletitle{OA-Mine: Open-World Attribute Mining for E-Commerce
  Products with Weak Supervision}. In \bibinfo{booktitle}{\emph{Proceedings of
  the ACM Web Conference 2022}} (Virtual Event, Lyon, France)
  \emph{(\bibinfo{series}{WWW '22})}. \bibinfo{publisher}{Association for
  Computing Machinery}, \bibinfo{address}{New York, NY, USA},
  \bibinfo{pages}{3153–3161}.
\newblock
\showISBNx{9781450390965}
\urldef\tempurl%
\url{https://doi.org/10.1145/3485447.3512035}
\showDOI{\tempurl}


\bibitem[Zhang et~al\mbox{.}(2023)]%
        {zhang2023pay}
\bibfield{author}{\bibinfo{person}{Yupeng Zhang}, \bibinfo{person}{Shensi
  Wang}, \bibinfo{person}{Peiguang Li}, \bibinfo{person}{Guanting Dong},
  \bibinfo{person}{Sirui Wang}, \bibinfo{person}{Yunsen Xian},
  \bibinfo{person}{Zhoujun Li}, {and} \bibinfo{person}{Hongzhi Zhang}.}
  \bibinfo{year}{2023}\natexlab{}.
\newblock \showarticletitle{Pay attention to implicit attribute values: a
  multi-modal generative framework for AVE task}. In
  \bibinfo{booktitle}{\emph{Findings of the Association for Computational
  Linguistics: ACL 2023}}. \bibinfo{pages}{13139--13151}.
\newblock


\bibitem[Zheng et~al\mbox{.}(2018)]%
        {10.1145/3219819.3219839}
\bibfield{author}{\bibinfo{person}{Guineng Zheng}, \bibinfo{person}{Subhabrata
  Mukherjee}, \bibinfo{person}{Xin~Luna Dong}, {and} \bibinfo{person}{Feifei
  Li}.} \bibinfo{year}{2018}\natexlab{}.
\newblock \showarticletitle{OpenTag: Open Attribute Value Extraction from
  Product Profiles}. In \bibinfo{booktitle}{\emph{Proceedings of the 24th ACM
  SIGKDD International Conference on Knowledge Discovery \&amp; Data Mining}}
  (London, United Kingdom) \emph{(\bibinfo{series}{KDD '18})}.
  \bibinfo{publisher}{Association for Computing Machinery},
  \bibinfo{address}{New York, NY, USA}, \bibinfo{pages}{1049–1058}.
\newblock
\showISBNx{9781450355520}
\urldef\tempurl%
\url{https://doi.org/10.1145/3219819.3219839}
\showDOI{\tempurl}


\bibitem[Zhu et~al\mbox{.}(2020)]%
        {zhu-etal-2020-multimodal}
\bibfield{author}{\bibinfo{person}{Tiangang Zhu}, \bibinfo{person}{Yue Wang},
  \bibinfo{person}{Haoran Li}, \bibinfo{person}{Youzheng Wu},
  \bibinfo{person}{Xiaodong He}, {and} \bibinfo{person}{Bowen Zhou}.}
  \bibinfo{year}{2020}\natexlab{}.
\newblock \showarticletitle{Multimodal Joint Attribute Prediction and Value
  Extraction for {E}-commerce Product}. In
  \bibinfo{booktitle}{\emph{Proceedings of the 2020 Conference on Empirical
  Methods in Natural Language Processing (EMNLP)}}.
  \bibinfo{publisher}{Association for Computational Linguistics},
  \bibinfo{address}{Online}, \bibinfo{pages}{2129--2139}.
\newblock
\urldef\tempurl%
\url{https://doi.org/10.18653/v1/2020.emnlp-main.166}
\showDOI{\tempurl}


\end{thebibliography}

%%
%% If your work has an appendix, this is the place to put it.
\section{Appendix}
\begin{table*}[]
\centering
\caption{Example of zero-shot dataset statistics in training, validation and testing sets, respectively.}
\label{tab:datasplits}
\begin{tabular}{l|ccc|ccc|ccc|c}
\hline
\multirow{2}{*}{Category} & \multicolumn{3}{c|}{Training} & \multicolumn{3}{c|}{Validation} & \multicolumn{3}{c|}{Testing} & All         \\ \cline{2-11} 
                          & \#P      & \#A     & \#PA     & \#P      & \#A      & \#PA      & \#P     & \#A     & \#PA     & Ave \#A/P \\ \hline
Arts                      & 10,250   & 1,796   & 8,400    & 3        & 6        & 6         & 15      & 23      & 30       & 2.48        \\
Books                     & 9,310    & 158     & 5,210    & 4        & 3        & 8         & 413     & 54      & 852      & 1.44        \\
Cellphones                & 6,772    & 1,149   & 5,187    & 91       & 109      & 192       & 157     & 175     & 332      & 2.38        \\
Giftcards                 & 84       & 8       & 74       & 8        & 2        & 16        & 11      & 3       & 9        & 2.37        \\
Grocery                   & 15,834   & 3,945   & 13,933   & 8        & 16       & 16        & 18      & 33      & 36       & 2.56        \\
Industrial                & 2,644    & 1,264   & 2,381    & 16       & 27       & 33        & 8       & 14      & 17       & 2.76        \\
Pet                       & 12,878   & 2,193   & 13,187   & 24       & 42       & 48        & 73      & 117     & 150      & 3.16        \\
Software                  & 187      & 87      & 152      & 2        & 4        & 4         & 8       & 14      & 16       & 2.11        \\
Tools                     & 30,236   & 6,210   & 29,759   & 14       & 24       & 28        & 58      & 97      & 120      & 2.92        \\
Videogames                & 559      & 240     & 477      & 35       & 45       & 75        & 57      & 67      & 128      & 2.86        \\ \hline
\end{tabular}
\end{table*}

\begin{table*}[!htp]
\small
\centering
\caption{Model Training Time in One Epoch (second).}
\label{tab:time}
\begin{tabular}{lcccccccccc}
\hline
Model      & Giftcards & Software & Videogames & Industrial & Cellphones & Arts    & Pet     & Books   & Grocery & Tools   \\ \hline
BERT-MLC   & 2.37      & 15.48    & 42.12 & 291.60     & 578.78     & 797.11  & 1004.53 & 1073.65 & 1266.68 & 2391.12 \\
BART       & 3.66      & 24.48    & 66.60 & 304.56     & 873.36     & 1152.00 & 1292.95 & 1604.52 & 1910.52 & 3521.88 \\
T5\textsubscript{small}     & 2.21      & 19.14    & 58.23 & 256.70     & 698.10     & 967.87  & 1209.27 & 1355.23 & 1576.67 & 2890.03 \\
GNN        & 0.09      & 0.19     & 1.00  & 5.46       & 16.46      & 30.02   & 57.50   & 12.80   & 73.33   & 150.91  \\
HGNN & 0.72      & 1.60     & 6.28  & 27.59      & 64.24      & 122.06  & 209.28  & 94.52   & 235.20  & 504.61  \\
\textbf{HyperPAVE}  & 0.90      & 1.66     & 6.71  & 30.06      & 70.18      & 133.43  & 224.40  & 89.22   & 251.14  & 543.07  \\ \hline
\end{tabular}
\end{table*}

\label{sec:appendix}
\subsection{Dataset}~\label{sec:appendix_dataset}
% Please add the following required packages to your document preamble:
% \usepackage{multirow}
% Please add the following required packages to your document preamble:
% \usepackage{multirow}

% Please add the following required packages to your document preamble:

Table~\ref{tab:datasplits} reports an example of dataset statistics in training, validation, and testing sets, where $\#P$, $\#A$, and $\#PA$ denotes the number of product nodes, the number of aspect nodes and the number of product to aspect edges, respectively.
The last column Ave $\#A/p$ indicates the average number of attribute value pairs for each product.
Because training, validation, and testing sets for the multi-label zero-shot setting are randomly generated for each run of the experiment, there exist different dataset statistics.

\subsection{Parameter Settings}~\label{sec:appendix_parameter}
We randomly select unseen attribute value pairs with unseen products following the sampling rule in Sec.~\ref{sec:zero_sampling}.
For the hyperparameter and configuration of HyperPAVE, we implement HyperPAVE in PyTorch and optimize it with AdamW optimizer. 
We train HyperPAVE and all baselines on the training set and we use a validation set to select the optimal hyper-parameter settings, and finally report the performance on the test set.
We follow the early stopping strategy when selecting the model for testing.
%The weights of the aspect node embeddings are 0.5 and 0.5 from `products with all aspects' and `category with all products and aspects' hyperedges.
For all methods, we run 10 times with different random seeds and report the average results with standard deviation.
Our proposed model HyperPAVE achieves its best performance with the following setup. 
The nodes' features are initialized by a BERT encoder with a 768-dimension size. The max length for category, product, and attribute values are 32, 512, and 32, respectively.
The initial learning rate is selected via grid search within the range of $\begin{Bmatrix} 5e-1, 5e-3, 5e-4, 5e-5 \end{Bmatrix}$ with 1e-6 weight decay for minimizing the loss.
The hidden sizes for convolution layers are 768 in both HyperConv and GraphConv. 
The activation function is ReLU.
The dropout rate is 0.5 and the batch size is 4.
We set the number of neighbors to 20 and the negative sampling rate is 2.0.
For the fusion module, the weights of the product node embeddings from hyperedges of `also buy', `also view', `products with all aspects', and `category with all products and aspects' are dynamically changed for different categories.
Experiments are conducted in Sec.~\ref{sec:sensitivity} to explore the weights in these fusion modules.

\subsection{Experiments}~\label{sec:more_experiments}
\begin{table*}[]
\small
\centering
\caption{Ablation study over HyperPAVE components in the zero-shot setting across seven categories on MAVE dataset.}
\label{tab:ablations}
\begin{tabular}{l|cccccccc}
\hline
                                    & F1                                         & mAP                                        & AUC                                        & MRR                                        & NDCG                                       & Hit@5                                      & Hits@10                                    & Hits@100                                    \\ \hline
                                    & \multicolumn{8}{c}{Arts}                                                                                                                                                                                                                                                                                                                                               \\ \hline
nodeID                              & 1.35 $\pm$ 0.29                            & 10.73 $\pm$ 0.52                           & 92.03 $\pm$ 0.12                           & 0.86 $\pm$ 0.03                            & 21.92 $\pm$ 0.58                           & 15.00 $\pm$ 1.23                           & 28.33 $\pm$ 1.10                           & 61.67 $\pm$ 0.64                            \\
BERT                                & 8.23 $\pm$ 0.24                            & 26.30 $\pm$ 0.14                           & 92.69 $\pm$ 0.06                           & 11.48 $\pm$ 0.19                           & 40.27 $\pm$ 0.15                           & 35.43 $\pm$ 0.27                           & 52.86 $\pm$ 0.14                           & 82.43 $\pm$ 0.10                            \\
BERT (Fine-tuned)                   & 19.87 $\pm$ 0.15                           & 34.77 $\pm$ 0.39                           & \textbf{99.84 $\pm$ 0.04} & 13.16 $\pm$ 0.17                           & 53.84 $\pm$ 0.25                           & 75.00 $\pm$ 0.00                           & 75.00 $\pm$ 0.00                           & 100.00 $\pm$ 0.00                           \\
Hyper (Product)                     & 30.93 $\pm$ 0.26                           & 42.33 $\pm$ 0.27                           & 99.06 $\pm$ 0.01                           & 22.95 $\pm$ 0.27                           & 57.84 $\pm$ 0.24                           & 57.50 $\pm$ 0.31                           & 75.00 $\pm$ 0.03                           & 93.75 $\pm$ 0.04                            \\
Hyper (Behavior)                    & 37.03 $\pm$ 0.67                           & 42.39 $\pm$ 0.62                           & 98.35 $\pm$ 0.02                           & 28.51 $\pm$ 0.50                           & 60.47 $\pm$ 0.46                           & 56.67 $\pm$ 0.29                           & \textbf{83.33 $\pm$ 0.24} & 100.00 $\pm$ 0.00                           \\
\textbf{HyperPAVE} & \textbf{43.33 $\pm$ 0.22} & \textbf{40.99 $\pm$ 0.18} & 99.22 $\pm$ 0.01                           & \textbf{47.52 $\pm$ 0.30} & \textbf{64.87 $\pm$ 0.39} & \textbf{75.00 $\pm$ 0.00} & 82.50 $\pm$ 0.12                           & \textbf{100.00 $\pm$ 0.00} \\ \hline
                                    & \multicolumn{8}{c}{Cellphones}                                                                                                                                                                                                                                                                                                                                         \\ \hline
nodeID                              & 19.75 $\pm$ 0.67                           & 22.88 $\pm$ 0.20                           & 97.72 $\pm$ 0.01                           & 10.65 $\pm$ 0.45                           & 38.33 $\pm$ 0.23                           & 38.33 $\pm$ 0.50                           & 57.22 $\pm$ 0.15                           & 80.00 $\pm$ 0.11                            \\
BERT                                & 24.21 $\pm$ 0.15                           & 26.15 $\pm$ 0.16                           & 97.60 $\pm$ 0.02                           & 17.02 $\pm$ 0.47                           & 40.97 $\pm$ 0.24                           & 41.11 $\pm$ 0.05                           & 70.05 $\pm$ 0.30                           & 86.11 $\pm$ 0.36                            \\
BERT (Fine-tuned)                   & 22.77 $\pm$ 0.16                           & 26.80 $\pm$ 0.31                           & 98.09 $\pm$ 0.04                           & 16.50 $\pm$ 0.43                           & 43.03 $\pm$ 0.37                           & 50.00 $\pm$ 0.00                           & 75.00 $\pm$ 0.00                           & 92.50 $\pm$ 0.12                            \\
Hyper (Product)                     & 32.27 $\pm$ 0.28                           & 33.32 $\pm$ 0.09                           & 98.94 $\pm$ 0.04                           & 22.26 $\pm$ 0.30                           & \textbf{54.17 $\pm$ 0.25}                  & 70.25 $\pm$ 0.27                           & \textbf{90.00 $\pm$ 0.00}                  & 100.00 $\pm$ 0.00                           \\
Hyper (Behavior)                    & 28.32 $\pm$ 0.38                           & 33.88 $\pm$ 0.22                           & \textbf{99.63 $\pm$ 0.01}                  & 22.57 $\pm$ 0.10                           & 47.81 $\pm$ 0.26                           & 52.50 $\pm$ 0.14                           & 61.67 $\pm$ 0.04                           & 97.50 $\pm$ 0.08                            \\
\textbf{HyperPAVE} & \textbf{39.91 $\pm$ 0.16}                  & \textbf{35.81 $\pm$ 0.18}                  & 99.22 $\pm$ 0.02                           & \textbf{23.54 $\pm$ 0.20}                  & 52.88 $\pm$ 0.18                           & \textbf{72.50 $\pm$ 0.08}                  & 75.00 $\pm$ 0.00                           & \textbf{100.00 $\pm$ 0.00}                  \\ \hline
                                    & \multicolumn{8}{c}{Grocery}                                                                                                                                                                                                                                                                                                                                            \\ \hline
nodeID                              & 6.50 $\pm$ 0.49                            & 23.31 $\pm$ 0.27                           & 95.48 $\pm$ 0.04                           & 15.33 $\pm$ 0.19                           & 21.98 $\pm$ 0.38                           & 22.50 $\pm$ 0.28                           & 35.00 $\pm$ 0.31                           & 65.00 $\pm$ 0.27                            \\
BERT                                & 14.65 $\pm$ 0.40                           & 22.85 $\pm$ 0.34                           & 96.18 $\pm$ 0.08                           & 15.80 $\pm$ 0.33                           & 22.55 $\pm$ 0.42                           & 30.10 $\pm$ 0.31                           & 35.10 $\pm$ 0.17                           & 75.00 $\pm$ 0.51                            \\
BERT (Fine-tuned)                   & 19.42 $\pm$ 0.46                           & 25.84 $\pm$ 0.18                           & 99.20 $\pm$ 0.01                           & 17.78 $\pm$ 0.20                           & 27.93 $\pm$ 0.29                           & 25.00 $\pm$ 0.10                           & 35.50 $\pm$ 0.13                           & \textbf{87.50 $\pm$ 0.13}                   \\
Hyper (Product)                     & 22.41 $\pm$ 0.62                           & 32.41 $\pm$ 0.37                           & 99.48 $\pm$ 0.02                           & 18.82 $\pm$ 0.28                           & 35.64 $\pm$ 0.40                           & 33.33 $\pm$ 0.71                           & 35.50 $\pm$ 0.21                           & 66.67 $\pm$ 0.10                            \\
Hyper (Behavior)                    & 29.20 $\pm$ 0.29                           & 32.85 $\pm$ 0.49                           & 98.34 $\pm$ 0.04                           & 14.41 $\pm$ 0.16                           & 37.66 $\pm$ 0.37                           & 35.05 $\pm$ 0.16                           & 50.00 $\pm$ 0.00                           & 70.00 $\pm$ 0.11                            \\
\textbf{HyperPAVE} & \textbf{33.43 $\pm$ 0.28}                  & \textbf{42.71 $\pm$ 0.30}                  & \textbf{99.56 $\pm$ 0.00}                  & \textbf{22.52 $\pm$ 0.38}                  & \textbf{52.64 $\pm$ 0.36}                  & \textbf{50.00 $\pm$ 0.00}                  & \textbf{50.00 $\pm$ 0.00}                  & 75.50 $\pm$ 0.50                            \\ \hline
                                    & \multicolumn{8}{c}{Industrial}                                                                                                                                                                                                                                                                                                                                         \\ \hline
nodeID                              & 10.40 $\pm$ 0.38                           & 16.44 $\pm$ 0.22                           & 93.16 $\pm$ 0.05                           & 2.59 $\pm$ 0.17                            & 30.07 $\pm$ 0.24                           & 28.75 $\pm$ 0.49                           & 35.00 $\pm$ 0.35                           & 68.75 $\pm$ 0.27                            \\
BERT                                & 1.48 $\pm$ 0.24                            & 5.37 $\pm$ 0.16                            & 89.75 $\pm$ 0.11                           & 0.66 $\pm$ 0.10                            & 13.58 $\pm$ 0.32                           & 8.13 $\pm$ 0.31                            & 11.87 $\pm$ 0.54                           & 55.63 $\pm$ 0.71                            \\
BERT (Fine-tuned)                   & 14.06 $\pm$ 0.11                           & 18.82 $\pm$ 0.50                           & 99.05 $\pm$ 0.01                           & 4.99 $\pm$ 0.14                            & 41.11 $\pm$ 0.50                           & 25.00 $\pm$ 0.00                           & 50.00 $\pm$ 0.04                           & 100.00 $\pm$ 0.00                           \\
Hyper (Product)                     & 19.78 $\pm$ 0.19                           & 14.15 $\pm$ 0.17                           & 94.34 $\pm$ 0.08                           & 7.63 $\pm$ 0.16                            & 26.68 $\pm$ 0.16                           & 24.73 $\pm$ 0.29                           & 37.50 $\pm$ 0.20                           & 75.00 $\pm$ 0.40                            \\
Hyper (Behavior)                    & 15.70 $\pm$ 0.31                           & 31.42 $\pm$ 0.30                           & 96.57 $\pm$ 0.04                           & 7.19 $\pm$ 0.33                            & 45.26 $\pm$ 0.22                           & 41.25 $\pm$ 0.31                           & 55.00 $\pm$ 0.35                           & 87.50 $\pm$ 0.00                            \\
\textbf{HyperPAVE} & \textbf{27.70 $\pm$ 0.10} & \textbf{33.29 $\pm$ 0.17} & \textbf{99.71 $\pm$ 0.01} & \textbf{16.10 $\pm$ 0.08} & \textbf{54.08 $\pm$ 0.26} & \textbf{52.50 $\pm$ 0.18} & \textbf{80.00 $\pm$ 0.16} & \textbf{100.00 $\pm$ 0.00} \\ \hline
                                    & \multicolumn{8}{c}{Software}                                                                                                                                                                                                                                                                                                                                           \\ \hline
nodeID                              & 1.97 $\pm$ 0.22                            & 18.11 $\pm$ 0.25                           & 76.27 $\pm$ 0.18                           & 4.39 $\pm$ 0.32                            & 30.12 $\pm$ 0.53                           & 23.75 $\pm$ 0.58                           & 62.50 $\pm$ 0.05                           & 100.00 $\pm$ 0.00                           \\
BERT                                & 7.38 $\pm$ 0.14                            & 14.10 $\pm$ 0.31                           & 74.89 $\pm$ 0.14                           & 6.38 $\pm$ 0.29                            & 34.19 $\pm$ 0.44                           & 26.70 $\pm$ 0.20                           & 36.25 $\pm$ 0.11                           & 100.00 $\pm$ 0.00                           \\
BERT (Fine-tuned)                   & 11.78 $\pm$ 0.31                           & 15.29 $\pm$ 0.50                           & 76.70 $\pm$ 0.03                           & 6.75 $\pm$ 0.46                            & 36.52 $\pm$  0.45                          & 23.75 $\pm$ 0.23                           & 37.50 $\pm$ 0.16                           & 100.00 $\pm$ 0.00                           \\
Hyper (Product)                     & 35.88 $\pm$ 0.37                           & 40.72 $\pm$ 0.16                           & \textbf{84.40 $\pm$ 0.10} & 21.25 $\pm$ 0.46                           & 59.51 $\pm$ 0.18                           & 46.25 $\pm$ 0.26                           & \textbf{63.75 $\pm$ 0.40} & 100.00 $\pm$ 0.00                           \\
Hyper (Behavior)                    & 12.22 $\pm$ 0.36                           & 36.33 $\pm$ 0.48                           & 81.25 $\pm$ 0.10                           & 6.09 $\pm$ 0.27                            & 34.19 $\pm$ 0.20                           & 25.00 $\pm$ 0.31                           & 38.75 $\pm$ 0.51                           & 100.00 $\pm$ 0.00                           \\
\textbf{HyperPAVE} & \textbf{47.62 $\pm$ 0.21} & \textbf{51.64 $\pm$ 0.10} & 77.80 $\pm$ 0.12                           & \textbf{26.66 $\pm$ 0.15} & \textbf{63.48 $\pm$ 0.10} & \textbf{61.25 $\pm$ 0.40} & 62.50 $\pm$ 0.25                           & \textbf{100.00 $\pm$ 0.00} \\ \hline
                                    & \multicolumn{8}{c}{Tools}                                                                                                                                                                                                                                                                                                                                              \\ \hline
nodeID                              & 8.90 $\pm$ 0.37                            & 17.00 $\pm$ 0.24                           & 97.91 $\pm$ 0.10                           & 2.36 $\pm$ 0.19                            & 22.27 $\pm$ 0.37                           & 50.00 $\pm$ 0.02                           & 50.00 $\pm$ 0.00                           & 50.00 $\pm$ 0.00                            \\
BERT                                & 14.53 $\pm$ 0.17                           & 18.51 $\pm$ 0.25                           & 96.21 $\pm$ 0.09                           & 6.51 $\pm$ 0.18                            & 21.30 $\pm$ 0.48                           & 48.50 $\pm$ 0.15                           & 52.05 $\pm$ 0.33                           & 80.00 $\pm$ 0.20                            \\
BERT (Fine-tuned)                   & 21.33 $\pm$ 0.14                           & 23.85 $\pm$ 0.36                           & 99.19 $\pm$ 0.05                           & 6.81 $\pm$ 0.31                            & 26.88 $\pm$ 0.32                           & 49.15 $\pm$ 0.26                           & 55.70 $\pm$ 0.39                           & \textbf{87.07 $\pm$ 0.25}                   \\
Hyper (Product)                     & 32.86 $\pm$ 0.24                           & 29.20 $\pm$ 0.47                           & 98.27 $\pm$ 0.06                           & 12.26 $\pm$ 0.14                           & 43.96 $\pm$ 0.26                           & 49.53 $\pm$ 0.35                           & 65.00 $\pm$ 0.24                           & 83.87 $\pm$ 0.17                            \\
Hyper (Behavior)                    & 31.43 $\pm$ 0.27                           & 25.13 $\pm$ 0.25                           & \textbf{99.30 $\pm$ 0.07}                  & 11.51 $\pm$ 0.17                           & 28.11 $\pm$ 0.23                           & 50.06 $\pm$ 0.25                           & 58.20 $\pm$ 0.34                           & 86.40 $\pm$ 0.18                            \\
\textbf{HyperPAVE} & \textbf{34.00 $\pm$ 0.28}                  & \textbf{47.83 $\pm$ 0.29}                  & 98.00 $\pm$ 0.06                           & \textbf{12.93 $\pm$ 0.18}                  & \textbf{59.05 $\pm$ 0.27}                  & \textbf{52.00 $\pm$ 0.34}                  & \textbf{65.37 $\pm$ 0.25}                  & 84.72 $\pm$ 0.20                            \\ \hline
                                    & \multicolumn{8}{c}{Videogames}                                                                                                                                                                                                                                                                                                                                         \\ \hline
nodeID                              & 3.25 $\pm$ 0.47                            & 7.31 $\pm$ 0.38                            & 79.00 $\pm$ 0.58                           & 1.49 $\pm$ 0.22                            & 17.27 $\pm$ 0.19                           & 10.00 $\pm$ 1.21                           & 20.00 $\pm$ 1.26                           & 70.00 $\pm$ 0.62                            \\
BERT                                & 6.67 $\pm$ 0.41                            & 10.25 $\pm$ 0.27                           & 85.83 $\pm$ 0.21                           & 3.01 $\pm$ 0.52                            & \textbf{33.30 $\pm$ 0.35}                  & 30.05 $\pm$ 0.26                           & 43.50 $\pm$ 0.35                           & 100.00 $\pm$ 0.00                           \\
BERT (Fine-tuned)                   & 12.87 $\pm$ 0.21                           & 11.44 $\pm$ 0.17                           & 76.84 $\pm$ 0.15                           & 4.21 $\pm$ 0.41                            & 25.26 $\pm$ 0.29                           & 15.71 $\pm$ 0.54                           & 37.86 $\pm$ 0.30                           & 73.70 $\pm$ 0.26                            \\
Hyper (Product)                     & 20.00 $\pm$ 0.23                           & 16.45 $\pm$ 0.19                           & 91.51 $\pm$ 0.20                           & 8.76 $\pm$ 0.39                            & 28.61 $\pm$ 0.25                           & 45.00 $\pm$ 0.16                           & 50.00 $\pm$ 0.15                           & \textbf{100.00 $\pm$ 0.00}                  \\
Hyper (Behavior)                    & 16.83 $\pm$ 0.26                           & 12.38 $\pm$ 0.11                           & 86.73 $\pm$ 0.36                           & 7.33 $\pm$ 0.16                            & 27.28 $\pm$ 0.17                           & 15.00 $\pm$ 0.55                           & 40.71 $\pm$ 0.38                           & 80.71 $\pm$ 0.35                            \\
\textbf{HyperPAVE} & \textbf{25.31 $\pm$ 0.19} & \textbf{21.19 $\pm$ 0.17}                  & 84.32 $\pm$ 0.05                           &\textbf{9.31 $\pm$ 0.30}  & 23.99 $\pm$ 0.16                           & \textbf{50.00 $\pm$ 0.50} & \textbf{50.00 $\pm$ 0.50}                  & 85.71 $\pm$ 0.12                            \\ \hline
\end{tabular}
\end{table*}

Due to limited space in the main context, we only demonstrate ablation study over three categories (Books, Giftcards, and Pets) in Table~\ref{tab:ablation}. Here in Table~\ref{tab:ablations}, we report the ablation study over the other seven categories on MAVE. We also demonstrate the model training time for one epoch across the ten categories on MAVE in Table~\ref{tab:time}. All models use the same input max\_length as 512 and batch size as 4. For different graph-based models, they show similar efficiency performance.
Thus, we only demonstrate two representative graph-based models (GNN and HGNN) for training efficiency comparison.

\end{document}